%% file: main.tex
\newif\ificlr
\iclrfalse

\documentclass[11pt]{article}

\ificlr
  \usepackage{iclr2027_conference,times}
  
  \def\paperbibliostyle{iclr2027_conference}
\else
  \usepackage{times}
  \usepackage{arxiv}
  \usepackage[numbers,sort&compress]{natbib}
  \usepackage{authblk}
  
  \def\paperbibliostyle{plainnat}
\fi

\input{math_commands.tex}

\usepackage{hyperref}
\usepackage{url}
\usepackage{graphicx}
\usepackage{cleveref}
\usepackage{tikz}
\usepackage{adjustbox}
\usetikzlibrary{shapes.geometric, arrows.meta, positioning, calc, fit, backgrounds, shadows}

\usepackage{xcolor}
\usepackage{booktabs}
\usepackage{multirow} 
\usepackage{amssymb}
\usepackage{algorithm}
\usepackage{algpseudocode}
\usepackage{subcaption}
\usepackage{placeins} 
\newcommand{\cmark}{\ensuremath{\checkmark}}
\newcommand{\xmark}{\ensuremath{\times}}

\definecolor{blockgreen}{RGB}{144, 190, 109}
\definecolor{gridline}{RGB}{40, 80, 30}
\definecolor{shapeblue}{RGB}{218, 228, 243}
\definecolor{badgegreen}{RGB}{210, 230, 205}
\definecolor{histblue}{RGB}{87, 132, 186}

\def\papertitle{Recycle only when you have to!}

\def\iclrAuthorBlock{%
Lorenz Vaitl$^{1\ast}$\thanks{Shared first authorship. Corresponding author: \texttt{lorenz\_maximilian.vaitl@roche.com}.},
Ivan Tanasijevic$^{1\ast}$,
Arian Rokkum Jamasb$^{1}$,
L.~Santiago Mille-Fragoso$^{2\dagger}$,
Pan Kessel$^{1}$\\
$^1$ Prescient Design, F. Hoffmann-La Roche Ltd.\\
$^2$ Stanford University\\
\ast\;Shared first authorship.\quad
\dagger\;Work done during an internship at Prescient Design, Roche.%
}

\def\arxivAuthorBlock{%
\author[1]{Lorenz Vaitl\thanks{Shared first author. Corresponding author: \texttt{lorenz\_maximilian.vaitl@roche.com}.}}
\author[1]{Ivan Tanasijevic\textsuperscript{*}}
\author[1]{Arian Rokkum Jamasb}
\author[2]{L.~Santiago Mille-Fragoso\thanks{Work done during an internship at Prescient Design, Roche.}}
\author[1]{Pan Kessel}

\affil[1]{Prescient Design, F. Hoffmann-La Roche Ltd.}
\affil[2]{Stanford University}%
}

\title{\papertitle}
\ificlr
  \author{\iclrAuthorBlock}
\else
  \arxivAuthorBlock
\fi

\begin{document}

\maketitle

\begin{abstract}
AlphaFold~3-like models (e.g., Boltz, Protenix, etc.) have pushed the frontier of protein binder design. This architecture is used both for the generation of binder candidates (e.g., by hallucination or converting these models into backbone or full atom generative models) as well as for the screening of these candidates for valid binders. In this paper, we start from the observation that AlphaFold~3's  confidence metrics (ipSAE, ipTM) as well as structural consistency metrics (scRMSD) can be predicted directly from the trunk \textit{even at early} recycling. Because our proposed predictor is cheap, differentiable, and available at every cycle, it can be
used in two distinct ways: as a decision signal to wisely spend the recycling budget for screening and
as an objective for hallucination-based design. We show that both applications lead to increased performance at significantly reduced computational cost.
\end{abstract}

\section{Introduction}

Several methods have recently succeeded in designing \textit{de novo} minibinders~\citep{rfdiffusion, colabfold, bindcraft, boltzgen} and  antibodies for specific epitopes~\citep{rfantibody, germinal, mber, boltzgen}. 
\textit{De novo} binder design typically involves a generative as well as a scoring step. The former creates plausible candidate protein binders, while the latter selects promising candidates for experimental validation. Our work improves both steps.

The screening step typically involves scoring the design by refolding its sequence. Designs are rejected if their refolds have low confidence or are structurally incompatible with the generative step. For state-of-the-art design pipelines, such as RFDiffusion and BindCraft, the computational cost is dominated by this refold. This is because the generative step of these pipelines involves an inverse folding that typically samples multiple sequences per design. As a concrete example, BindCraft typically uses 20 sequences per design. All of these sequences need to be refolded. The screening step has been shown to provide a significant boost to the binding rate~\citep{mber, bindcraft, rfantibody}. Furthermore, recent work suggests that sampling more candidates followed by more stringent screening also improves binding \citep{biohub2026languagemodeling}. A significant reduction of the numerical cost at the same refold fidelity is thus highly desirable.

The generative step can be implemented in different ways. A prominent approach involves hallucination~\cite{colabfold, bindcraft} for which the designed sequence is obtained by backpropagation through a folding model. This optimization is often quite unstable, which is unsurprising given that it involves large computational graphs. The size of the computational graph arises because the forward pass of the folding model involves recycles as well as a diffusion rollout. Finding ways to reduce the size of the computational graph at the same level of fidelity can thus be expected to improve the design quality as well as speed.  

\begin{figure}
    \centering
    \resizebox{\textwidth}{!}{
    \input{poppy-CC-scheme}
    }
    \caption{\textbf{CycleCritic for efficient screening.}
    After each recycle, CycleCritic predicts final-depth quality
    from the current trunk representations. It cancels unpromising
    refolds before the structure and confidence modules, avoiding
    further recycling and downstream evaluation. Retained candidates
    complete the full refold and are accepted only if they satisfy
    the final screening filters.}    \label{fig:visual-abstract}
\end{figure}
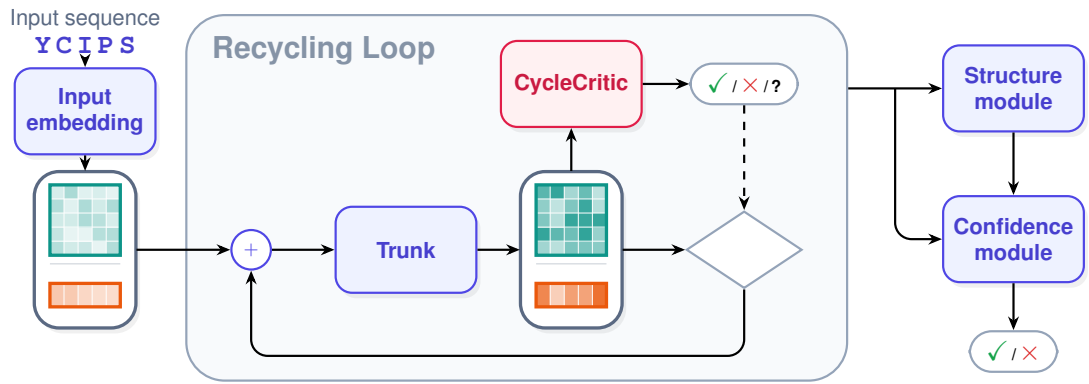

State-of-the-art folding models are based on the AlphaFold~3 architecture~\cite{abramson2024accurate}, retaining the fundamental recipe of MSA embedding, the recycled trunk, diffusion and confidence head. In recent hallucination pipelines, the AlphaFold~3 architecture is used in both the generative and scoring steps~\cite{biohub2026languagemodeling, boltzdesign}.

Our work starts from the observation that confidence as well as structural consistency can be predicted directly from the trunk even in an early recycle. Specifically, we demonstrate that one can train a lightweight \emph{CycleCritic} that can predict these quantities at end of each recycle from the pair representation. We then harness this predictor for both the generative and screening steps, achieving material improvements in compute allocation and binder hallucination in our experiments.

Our experiments demonstrate the benefits of the CycleCritic for screening. We generate $126{,}000$ VHH designs against seven antigens. We then show that the CycleCritic allows us to identify designs that pass both the confidence and self-consistency thresholds at a drastically reduced compute budget. Impressively, this is achieved at very low false positive rate. We also show that the CycleCritic outperforms all state-of-the-art baselines. For example, the CycleCritic is able to identify roughly $2.3$x more high quality designs than Protenix-mini per GPU hour. Impressively, it does so with 15 percentage point higher precision. fThe advantage is particularly pronounced for hard targets.

For the generative step, we apply the CycleCritic to two different hallucination approaches (ESMFold2 and Mosaic). We design for seven targets with both the VHH and scFV format resulting in a total of 14 design tasks. Our approach improves the top-50 designs on $13/14$ and $10/14$ design tasks for Mosaic and ESMFold respectively. 

Briefly, our main contributions are:
\begin{itemize}
    \item We analyse the dynamics of AF3-like models during recycling, show that the trunk is the main determining factor in folding performance which can often be predicted at early recycles.
    \item We propose a CycleCritic that can directly predict confidence and structural compatibility from the pair representation even at early recycle.
    \item For screening, we deploy the CycleCritic for dynamic allocation of the recycles. This drastically reduces the cost of passing designs since the model no longer needs to spend unnecessary recycles on designs that are easy to score.
    \item For hallucination-based generation, we use this CycleCritic to construct more economical confidence losses. We demonstrate that this leads to better design outcomes.
\end{itemize}

\section{Related Work}
\label{sec:related}
\paragraph{Interpreting the trunk representation.} \citet{migliorini2026pairsae} extract interpretable structural concepts directly from the pair representation of a co-folding model. Similarly, \citet{feldman2026alphainterp} probe AlphaFold~3's internal representations across the forward pass and find that the Pairformer compresses its inputs into a compact latent geometry in which complex biophysical features become linearly decodable.  \citet{lin2023evolutionary} and
\citet{biohub2026languagemodeling} investigate the ESMFold latent
space and show that real structural information is carried in the pair
representation. These works provide the foundation for our work by pointing out the richness of the trunk representation. 

\paragraph{Screening with trunk representation.} \citet{verburgt2025alphafold} refine AlphaFold's pLDDT head with equivariant graph neural networks over the predicted $\text{C}_\alpha$ coordinates emitted by the structure module.  \citet{liu2025af3score} adapt AlphaFold~3 to a score-only model by feeding input coordinates directly into the confidence head, bypassing thus the diffusion structure module. Both approaches require a structure to exist, while our method can predict confidences without a given structure. \citep{biohub2026languagemodeling} proposes a distogram ipTM 
proxy for ESMFold2. It has no learned parameters and provides a valuable baseline for our approach.

\paragraph{Generation with trunk representation.}
\citet{liu2026helixfolds1} derive inter-chain contact probabilities from the Pairformer trunk and use them as a blueprint constraining generation before
diffusion, pruning docking modes and reducing sampling roughly tenfold; they further note that these probabilities indicate prediction difficulty and sampling utility. ESMFold-based binder design~\cite{biohub2026languagemodeling} and
EasyNano~\citep{hu2026easynano} optimize against trunk outputs alone, computing their objective from the distogram or pair representation with no structure module in
the loop. Our contribution work extends on this body of literature by not only skipping diffusion but also dynamically allocating the recycle budget of the trunk.
\citet{jang2026synthesizing} improve folding models by learning to transfer the trunk representations of different models to each other.

\paragraph{Scoring and pose selection.} Confidence-based pose selection is a standard filter applied in co-folding-based workflows to improve hit-rates despite well-characterised limitations \citep{fromm2026eval, abagxm2026}. Recent works have focussed on developing structure-based antibody pose scorers trained with decoys. \citet{Xu2026deeprank} train an equivariant GNN for scoring antibody-antigen poses derived from classical docking methods. \citet{Grandguillaume2026arid} learn a scoring function from pLM embeddings, and force field and structure-derived features over docking decoys. \citet{Tadiello2026abag} train a Deep Set ranking model over an ensemble of AlphaFold~3 co-folding samples.

\section{Motivation: Dynamics of Recycling}
\label{sec:Dynamics}
\begin{figure}[ht]
    \centering
    \begin{subfigure}[t]{0.6\textwidth}
        \vspace{0pt}
        \centering
        \begin{tikzpicture}
            \node[inner sep=0, anchor=south west] (img1) {\includegraphics[width=\linewidth]{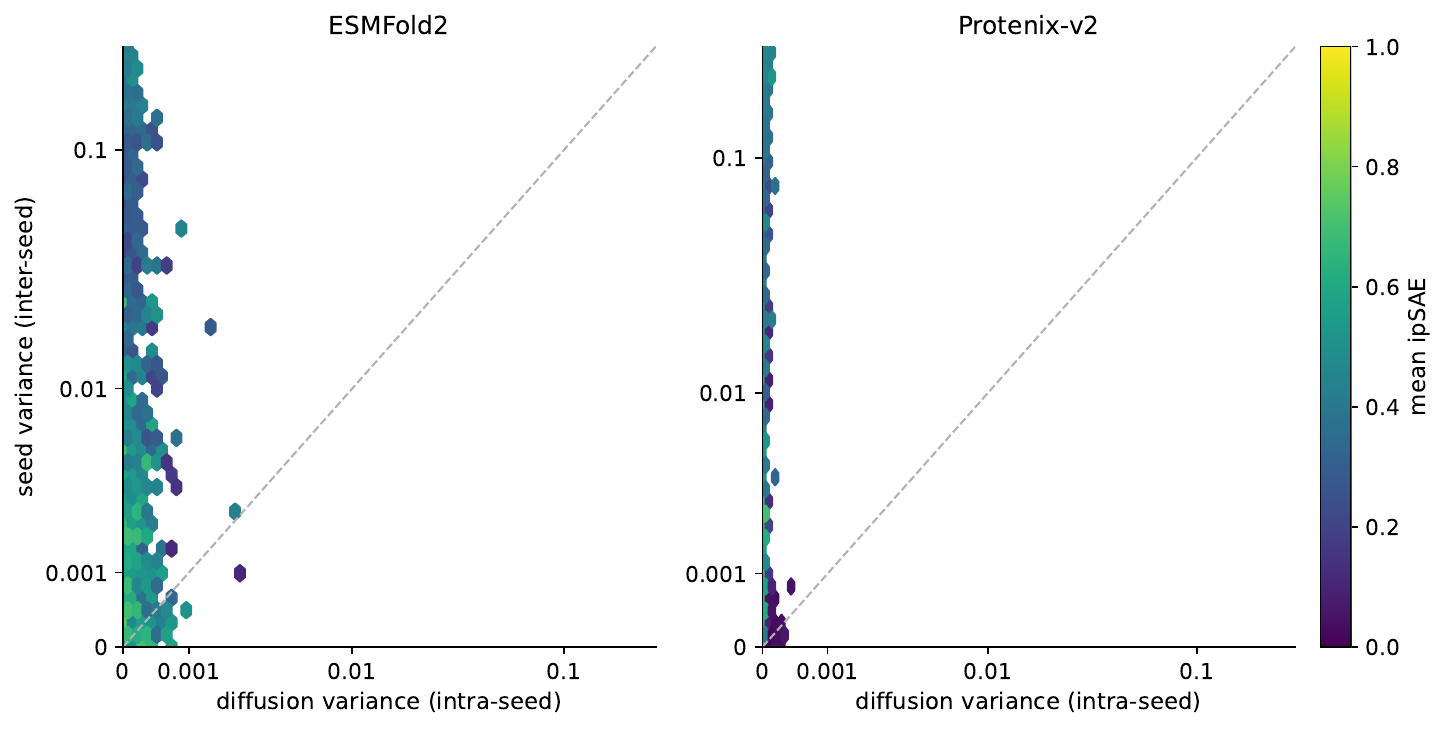}};
            \node[anchor=north west, font=\bfseries, xshift=2pt, yshift=-2pt] at (img1.north west) {a)};
        \end{tikzpicture}
        \phantomcaption\label{fig:variance}
    \end{subfigure}%
    \hfill
    \begin{subfigure}[t]{0.33\textwidth}
        \vspace{0pt}
        \centering
        \begin{tikzpicture}
            \node[inner sep=0, anchor=south west] (img2) {\includegraphics[width=\linewidth]{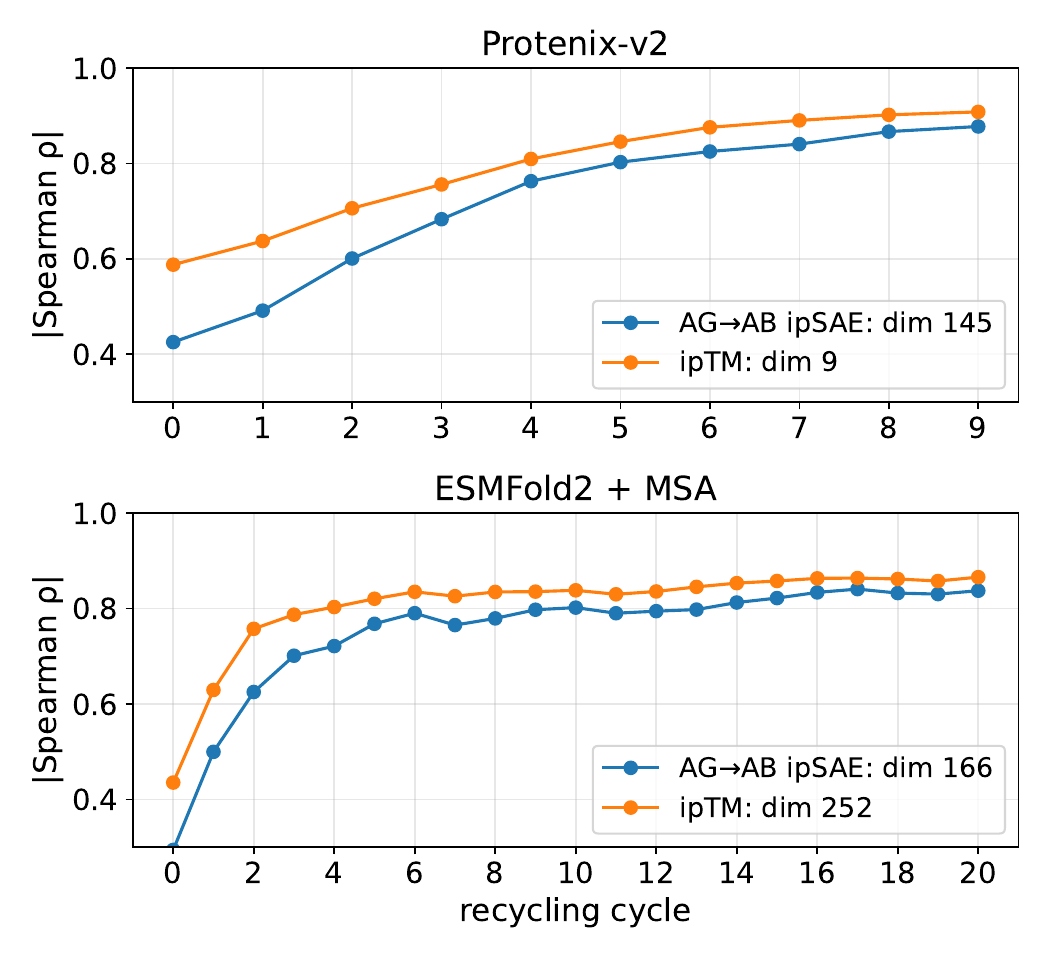}};
            \node[anchor=north west, font=\bfseries, xshift=-10pt, yshift=-2pt] at (img2.north west) {b)};
        \end{tikzpicture}
        \phantomcaption\label{fig:latent-dim-predictiveness}
    \end{subfigure}

    \vspace{0.5em}
    \caption{\textbf{Model variance and latent dimension predictability.} 
    \textbf{a) Seed and diffusion variance.} How much does the ipSAE vary between different seeds and diffusion samples. Computed with 3 seeds and diffusion samples each. 
    \textbf{b) Latent dimensions correlate with final confidence metrics.} We average-pool the latent dimensions over all pairs in the pair representations and take the highest correlated dimension. We see a high Spearman correlation with the final metric outputs even at early recycles. a) 2,500 b) 350 samples; both on Sabdab.}
    \label{fig:combined-dynamics}
\end{figure}

AlphaFold3-type folding models are generative models and therefore stochastic. We recall this stochasticity is not only because of the diffusion module but also due to the trunk. Specifically, recycles in the trunk are stochastic due to MSA subsampling and possible inference dropout. Our work is motivated by the observation that most of the variance of state-of-the-art AlphaFold~3-like models can be attributed to the trunk and not, as one may naively expect, the diffusion (Figure~\ref{fig:variance}).  We further observe that some components of the pair representation correlate closely with confidence metrics, even early during recycling, see Figure \ref{fig:latent-dim-predictiveness}.

\begin{figure}
    \centering
    \includegraphics[width=\linewidth]{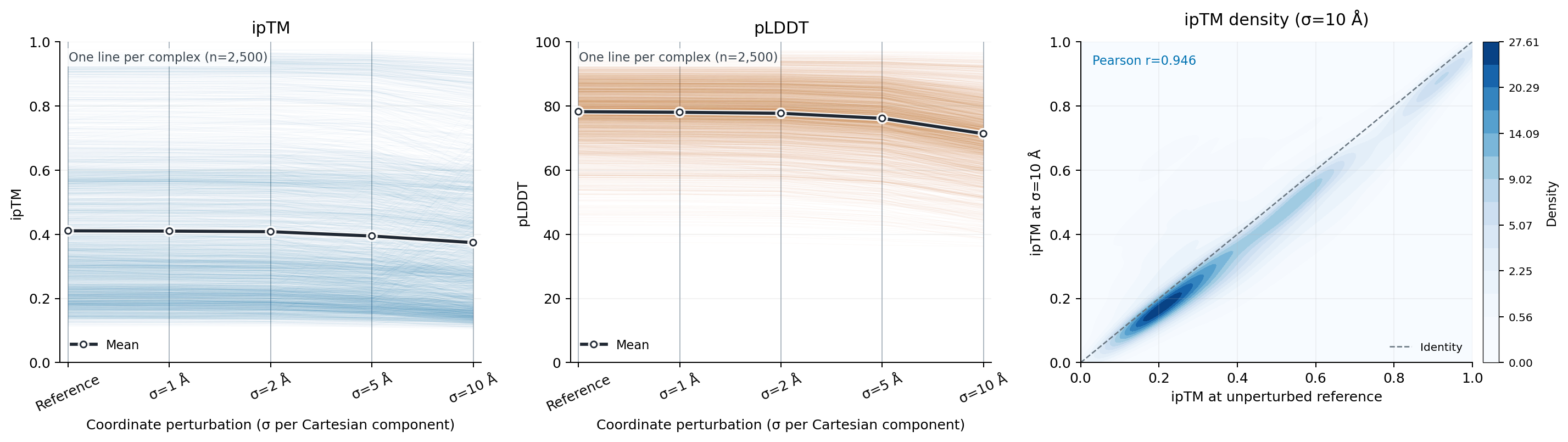}
    \caption{\textbf{Protenix-v2 Confidence metrics are largely determined by trunk representations.} Trunk representations were fixed while the confidence head scored perturbed structures for 2,500 antibody–antigen complexes randomly sampled from SAbDab. Perturbations comprised independent Gaussian noise with per-coordinate standard deviation \(\sigma \in \{1,2,5,10\}\,\text{Å}\). \textbf{Left:} ipTM across perturbations; traces denote individual complexes and black line denotes mean. \textbf{Centre:} pLDDT across perturbations; traces denote individual complexes and black line denotes mean. \textbf{Right:} Unperturbed versus perturbed ipTM for \(\sigma=10\,\text{Å}\); dashed diagonal indicates perfect agreement.}
    \label{fig:coord-perturb}
\end{figure}

To assess the dependence of confidence on the supplied structure, we fix trunk outputs and re-evaluate the Protenix-v2 and ESMFold2 confidence heads on perturbed coordinates for 2,500 antibody–antigen complexes. For Protenix-v2, adding Gaussian noise with $\sigma=$\(10\,\text{Å}\) modestly reduces ipTM and plDDT, with strong correlation between perturbed and unperturbed scores (Pearson \(r=0.946\)) (Figure \ref{fig:coord-perturb}). Input ablations localize the dominant contributor to confidence scores to the trunk’s pair representation. Removing the pair input collapses ipTM, while removing other inputs has limited effect (Figure \ref{fig:px-coordinate-perturbations}). Together, these observations indicate that confidence metrics are largely determined by the pair representation and only weakly adjusted by the supplied coordinates. We observe different, co-ordinate dependent, behaviour for ESMFold2, discussed in Appendix \ref{app:recycling-dynamics}. 

\section{CycleCritic}
The analysis of the previous section strongly suggests that one should be able to predict confidences as well as structural consistency directly from the pair representation even at early recycles. This then allows to dynamically allocate the recycling budget with the CycleCritic deciding the number of recyclings. In this section, we will outline our concrete implementation of this.

We use a simple DeepSets-style architecture \citep{zaheer2017deep}.
The architecture then consists of a shared per-pair encoder (no cross-attention), a permutation-invariant weighted pooling step, and
a small head on the pooled vector. We refer to Figure~\ref{fig:cyclecritic-combined} for an overview of the architecture.
\input{CC_combined_arch}

For each recycle, the CycleCritic is trained to predict the final metric after a fixed number of reference recycles. The number of reference recycles is chosen for each folding model such that it leads to high folding fidelity. This is motivated by the observation that prediction quality typically plateaus for a sufficiently large number of recycles. 

Self-consistency can only be predicted with respect to a given reference design structure while for confidence only the sequence of the design is sufficient. As a result, we train slightly different CycleCritic for structural consistency and confidence.

\paragraph{Confidence CycleCritic.} This critic takes the pair representation $z_{ij}^{(c)}$ of the current recycle as input as well as the recycle number $c$ and chain-identity. It then predicts the ipSAE, ipTM, pTM or pLDDT of the reference refold. The final head emits a categorical distribution $P_c \in \Delta^K$ over $K$ bins
tiling $[0,1]$, with scalar prediction $s_c = \sum_k P_{c,k}\, b_k$ for bin
centres $b_k$.

\paragraph{Structure CycleCritic.} This critic takes the pair representation $z_{ij}^{(c)}$ of the current recycle, the recycle number $c$, and the C$_\beta$ distance matrix of the design as input. Instead of directly using the pair representation for the per-pair embeddings, we found it advantageous to first transform it into the distogram with the frozen small distogram head. This head is part of the standard AlphaFold3 architecture and is used throughout the folding model training for the distogram loss. The resulting per-pair embeddings are then processed by the DeepSet-style architecture outlined above. The final head predicts the scRMSD discretized by 40 bins uniformly covering 0-20~\AA{}.

\subsection{Analytic Alternative}
We also propose analytic CycleCritics that require no learning, lowering the barrier to adoption at the cost of reduced performance. 

\paragraph{Structure CycleCritic.} We propose to leverage the distogram $p^{c}_{ij}$ predicted at cycle $c$ directly. We then compute the mean negative log likelihood of the design's distance matrix $D_{ij}$ across the binder-target interface, defined by $C_\alpha<12 \text{\AA}$ on the designed backbone. We refer to Algorithm~\ref{alg:nll-iface} for complete details. The log likelihood thresholds corresponding to the corresponding scRMSD thresholds can then be calibrated on a validation set. Alternatively, one can impose a log likelihood screening threshold directly.

\paragraph{Confidence CycleCritic.} For ipTM, a proxy was proposed in \citet{biohub2026languagemodeling} (Algorithm 15). Its basic idea is to use the entropy of the interface distogram as a proxy for ipTM. This is well motivated as the structure module produces the residue positions conditioned on the trunk. The distogram encodes how constraining this conditioning information is. Highly constrained structures come with lower uncertainty. We propose to use this proxy as an analytic Confidence CycleCritic. A downside of this approach is that it is restricted to ipTM.

As we discuss in Appendix~\ref{app:analytic_baseline}, these analytic alternatives are underperforming relative to their learned counterparts but still show meaningful performance.

\section{Experiments}
In the following, we demonstrate the power of the CycleCritic for screening and hallucination-based design. For both, we generate designs for seven benchmark targets: CD45, CTLA-4, EGFR, IL7Ra, INSR, PDGFR$\beta$, and PD-L1.

\subsection{Efficient Screening with CycleCritic}
\label{sec:Gating}

\begin{figure}[!htbp]

    \includegraphics[width=\linewidth]{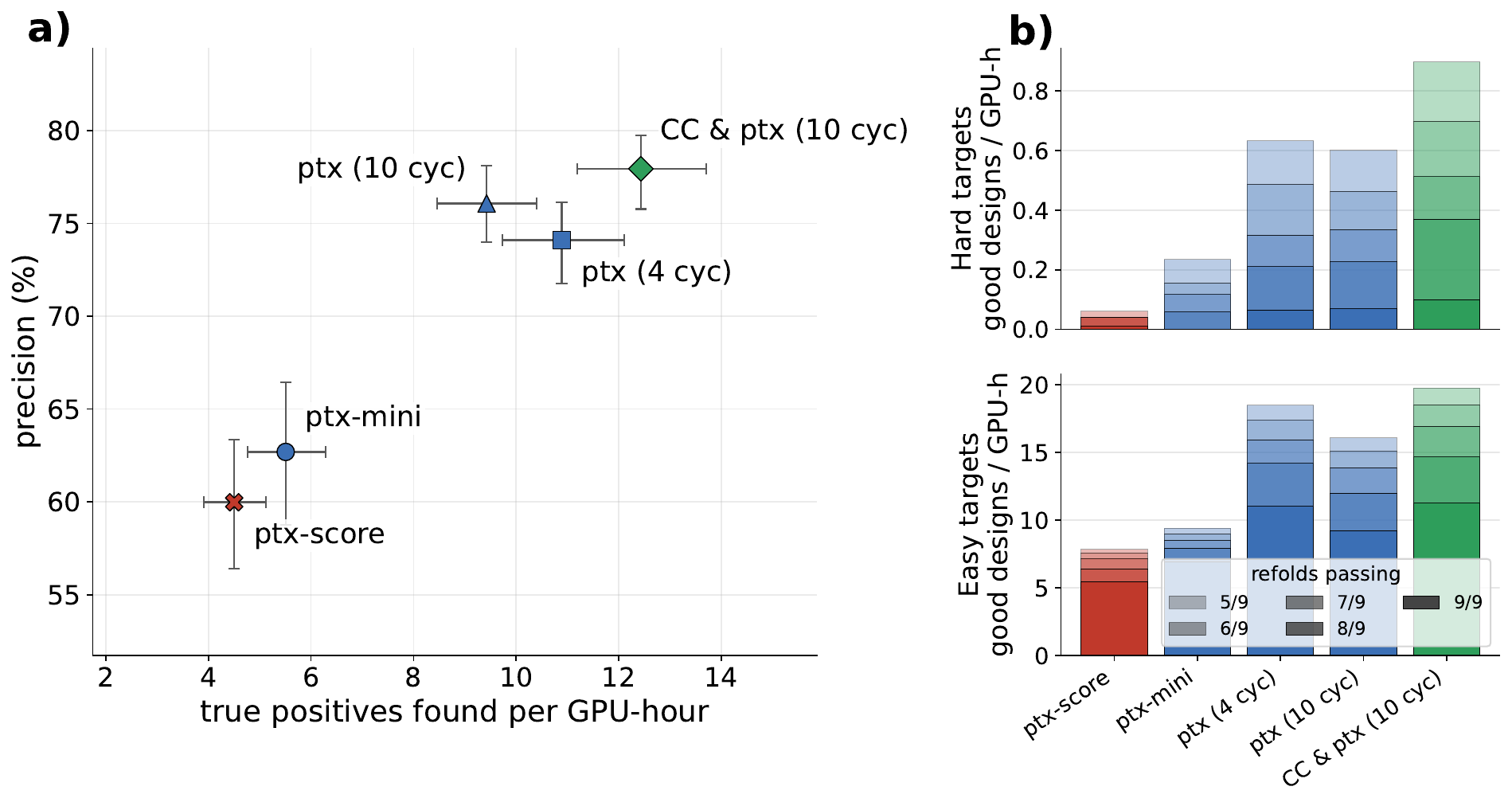}
    \caption{\textbf{CycleCritic outperforms all baselines.} 
    \textbf{a)} Precision and true positive throughput. 
    Error bars represent 95\% percentile bootstrap confidence intervals (2.5th--97.5th percentiles) computed across 2,000 draws by resampling 50-design chunks with replacement within each target.
    \textbf{b)} Hard vs. rest mean yield. 
Pooled mean good-design yield per GPU-hour across hard targets ($<100$ positives: CD45, CTLA-4, PDGFR$\beta$) and remaining targets (EGFR, IL7RA, PD-L1, INSR). Stacks show $5–9/9$ labels. 
    }
    \label{fig:tp-fp-per-gpu-hour}
\end{figure}

The CycleCritic can significantly speed up screening. To demonstrate this, we generated 126{,}000 VHH designs using BoltzGen \cite{stark2025boltzgen}. We use the seven benchmark targets described above and four VHH frameworks with PDB IDs \texttt{8coh}, \texttt{8z8v}, \texttt{7eow}, and \texttt{7xl0}. This yields 4500 designs per target--framework pair and $18{,}000$ designs per target.

These designs are then filtered. Protenix-v2 with 10 recycles is used as a reference, which is one of the leading folding models for antibody-antigen complexes~\cite{zhang2026protenixv2}. We aim to reproduce the fidelity of Protenix-v2 with 10 recycles at significantly lower computational cost.

We compare with three relevant baselines:
\begin{itemize}
    \item Protenix (4 cycles) \cite{chen2025protenix}: the same Protenix-v2 model with 4 instead of 10 recycles.
    \item Protenix-mini \cite{protenixmini}: a lightweight version of Protenix that is optimized for high inference speed at the cost of reduced folding fidelity.
    \item ProtenixScore \cite{protenixscore}: an implementation of the AF3Score architecture~\cite{liu2025af3score} for Protenix. This approach turns Protenix into a scorer. The trunk is templated on the input structure and the diffusion rollout is skipped. The confidence head takes as input the trunk embeddings as well as the design structure and produces the resulting confidence score. Note that this model cannot evaluate structural self-consistency.
\end{itemize}

Since AlphaFold3-like folding models are stochastic, we refold the same complex 9 times with Protenix-v2 at 10 recycles. Positives are designs that pass more than half the refolds with the thresholds $\text{ipSAE}>0.5$ and $\text{scRMSD}<3\text{\AA}$.
We then fold the design with all methods, check whether it passes the filters, and compare to the label obtained by the reference method. 

Ideally, a method can identify a high number of positives at drastically reduced runtime. Crucially, this needs to be achieved at high precision, i.e. the number of false positives needs to be low. Note that even the reference method will not perfectly agree with the reference labels due to the stochasticity of the refold. The independent repeat still serves as a reference for label agreement and screening performance. We quantify this by refolding the designs one more time with the reference method.

Fig.~\ref{fig:tp-fp-per-gpu-hour} demonstrates that the CycleCritic significantly outperforms all baselines. Impressively, it also matches the precision of the reference method within error bars. The CycleCritic is particularly beneficial for harder targets. We note that its performance can likely be improved by a more involved calibration approach. For the experiments, we prioritize convenience over maximal performance by adopting a very simple calibration method. Specifically, we chose the validation set of the cycle critic and adjusted the gating thresholds of the CycleCritic such that full recall is achieved, i.e., all positives in this set are recovered. We refer to Appendix~\ref{app:gating_campaign} for more details on the gating experiments.

\subsection{Hallucination-based Design with CycleCritic}
\label{sec:Hallucination}
In hallucination-based design~\cite{bindcraft, colabfold}, the amino acid sequence of the designed binder is optimized by gradient descent through the folding model. The optimization objective typically involves terms that try to maximize the confidence of the design.
For this reason, CycleCritic can be used in these terms to effectively approximate the confidence scores that would be obtained from a large number of recycles. In contrast, previous methods often use a low number of recycles or even detach some of them from the computational graph. We demonstrate that the CycleCritic approach improves antibody design for two state-of-the-art frameworks: Mosaic~\citep{boyd2025mosaic} with Protenix-v2 and the design strategy of ESMFold2~\citep{biohub2026languagemodeling}. We design antibodies both VHH and scFvs format for the seven benchmark targets. 

Interestingly, we observe better performance when averaging the CycleCritic-predicted terms across recycling loops, indicating that the CycleCritic provides useful information even in early recycling loops. 

\begin{figure}
    \centering
    \includegraphics[width=\linewidth]{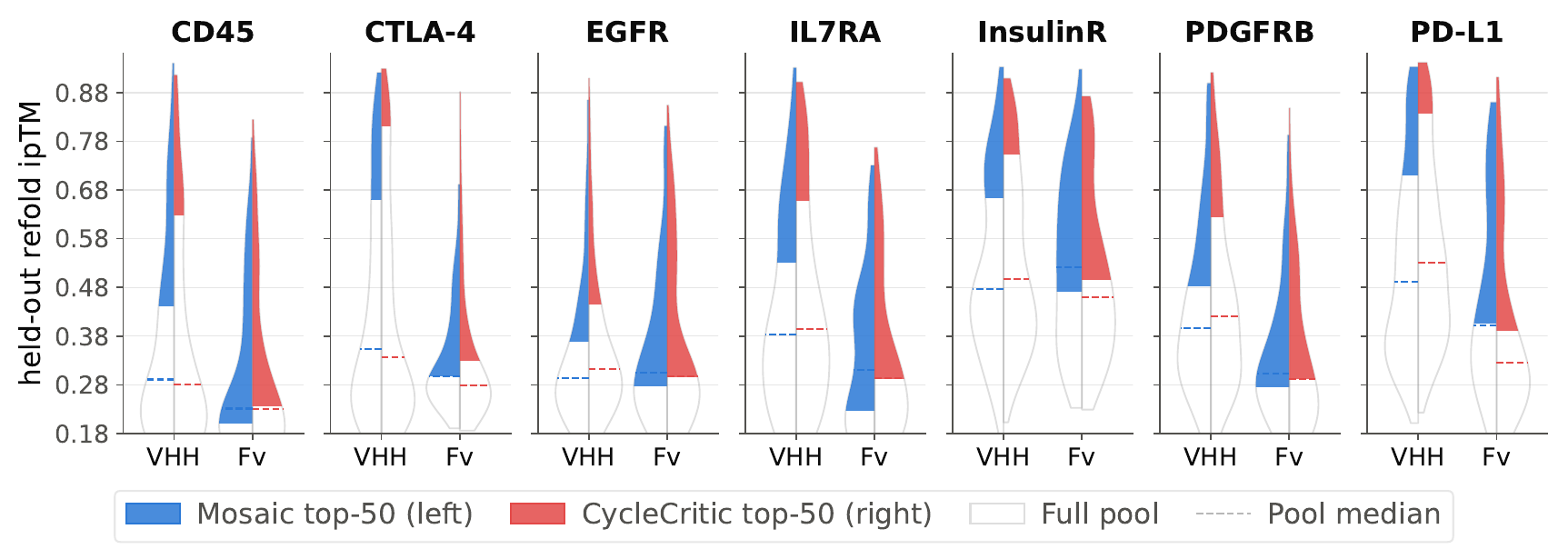}
    \caption{\textbf{For Mosaic, CycleCritic significantly improves top-$50$ designs.} Filled areas denote top-50 designs by refold ipTM and dashed lines are median over all produced designs. Both baseline and CycleCritic used the same 6 B200 compute hour budget per design task.}
\label{fig:Mosaic-metric}
\end{figure}

\paragraph{Mosaic.} We replace all confidence-head-based losses with alternatives derived from the CycleCritic. This allows us to skip the diffusion and confidence modules, generating nearly 60\% more samples within the same wall-clock time. We also replaced the $\text{pAE}$-based losses with analogous distogram-based losses.
For fair comparison, the final evaluations for both methods were performed using Protenix-v2 at 10 recycles for the refold. 

In protein design, one is typically only interested in the distribution over the best designs since wet-lab capacity is scarce. We therefore also compare the design strategies on the top-50 designs. Figure~\ref{fig:Mosaic-metric} demonstrates that the CycleCritic leads to substantial improvements. The top-50 mean is improved for 13 out of 14 design tasks (7 targets and 2 formats) with a mean gain of 6 p.p.. We refer to Appendix~\ref{App:Mosaic} for more details.

\begin{figure}
    \centering
    \includegraphics[width=\linewidth]{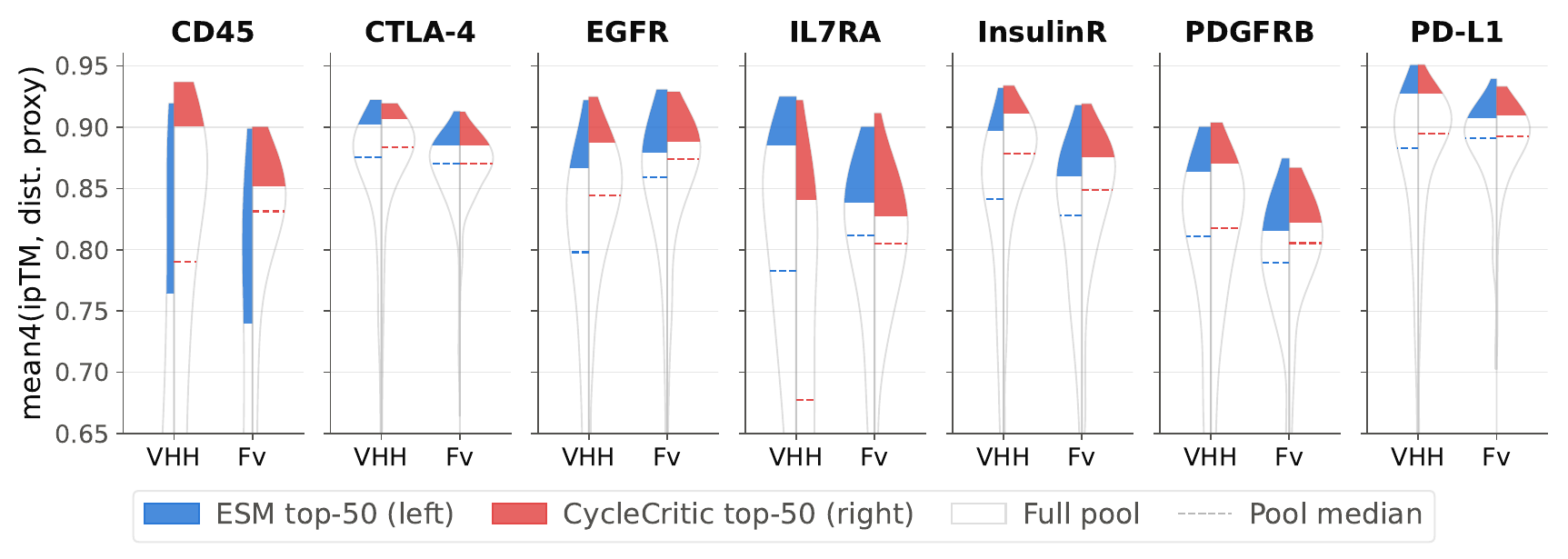}
    \caption{\textbf{For ESMFold2, CycleCritic raises per-design quality at matched
    compute.} We use the same ranking metric as in the ESMFold2 publication for comparison, i.e. the mean over ipTM and distogram-proxy ipTM averaged over 4 refolds with different checkpoints. Filled area is top-50 subset by the ranking metric. Dashed lines are medians over all produced designs.}
\label{fig:ESM-metric}
\end{figure}
\paragraph{ESMFold2.} This approach already skips the diffusion and confidence heads during hallucination, as it relies on distogram-based losses. Nevertheless, our experiments establish that optimization can strongly benefit from the CycleCritic. We attribute this to CycleCritic's ability to effectively predict confidence at large number of recycles, allowing us to estimate design confidence at higher fidelity at essentially the same computational cost. 

We add to the baseline optimization objective two terms based on the predicted ipTM and $\mathrm{ipSAE}_{\mathrm{AG}\rightarrow\mathrm{AB}}$.
We refer to Appendix~\ref{app:ESM-design} for more details.

We compare with the baseline for the same compute budget. Figure~\ref{fig:ESM-metric} demonstrates that the CycleCritic meaningfully improves over the baseline. Specifically, the CycleCritic improves the best designs in 10 out of the 14 design tasks by 1\%.  As was to be expected, performance can vary by target. CD45 sees a drastic improvement while IL7Ra leads to a small reduction.

\begin{figure}[t]
    \centering
    \includegraphics[width=0.85\linewidth]{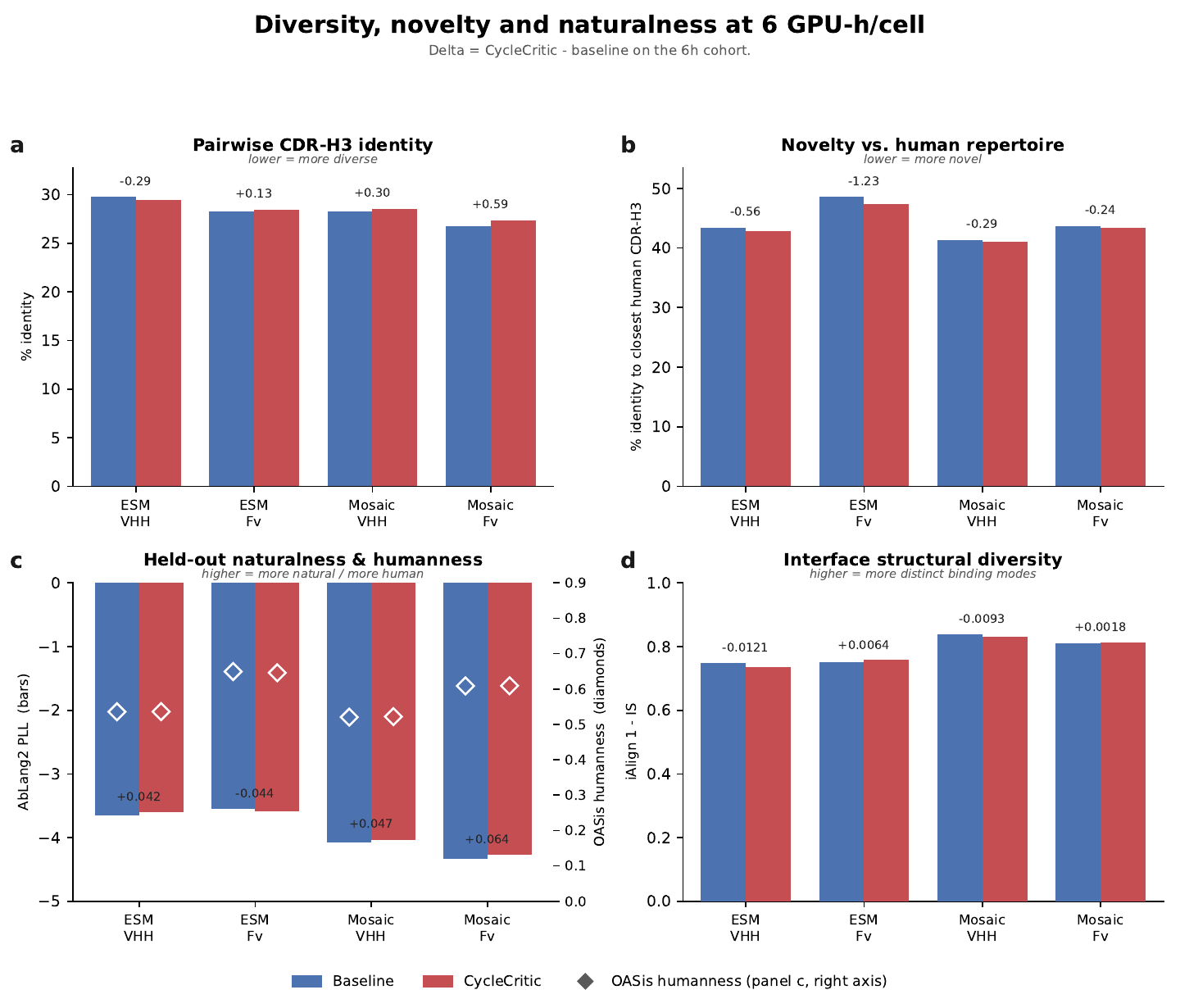}
    \caption{\textbf{Diversity, novelty and naturalness of designed binders.}
    Baseline (blue) versus CycleCritic (red) on the six-hour design
    cohorts. Annotations show CycleCritic minus baseline.
    \textbf{(a)} Mean pairwise CDR-H3 identity within
    target--framework cells.
    \textbf{(b)} Identity to the closest human CDR-H3 among
    $2.3$M unique heavy-chain loops from OAS.
    \textbf{(c)} AbLang2 pseudo-log-likelihood (bars, left axis)
    and OASis humanness (diamonds, right axis) over designed
    CDR positions.
    \textbf{(d)} Interface structural diversity measured by
    iAlign $1-\mathrm{IS}$ within target--framework--arm cells.
    The iAlign comparison describes the available structural
    subset, whose coverage differs between arms, rather than
    the complete design cohorts.}    \label{fig:CC-design}
\end{figure}

\paragraph{Design Quality.} To test whether CycleCritic degrades or collapses the designs, we measure a range of diversity and quality metrics. For both ESMFold2- and Mosaic-based design pipelines, CycleCritic designs mirror the diversity and novelty of their baselines. As summarized in \Cref{fig:CC-design}, we observe comparable CDR diversity, novelty with respect to human repertoire, humanness, and interface diversity. More details can be found in the Appendix~\ref{app:analysis_campaign}.


\FloatBarrier
\section{Conclusion}
A simple CycleCritic uses trunk representations from early recycles
to predict the confidence and structural consistency metrics that
would be obtained after a large number of recycles. This can speed
up and improve both the design and screening steps of binder design
pipelines. Specifically, the CycleCritic allows to dynamically
allocate recycles depending on the difficulty of the protein complex
at hand. This is somewhat reminiscent of reasoning where compute
budget is dynamically allocated by a language model.


\bibliography{iclr2027_conference}
\bibliographystyle{\paperbibliostyle}

\newpage
\appendix
\numberwithin{figure}{section}
\numberwithin{table}{section}

\section{Dynamics during recycling}
\label{app:recycling-dynamics}
\begin{figure}[h]
    \centering
    \includegraphics[width=0.8\linewidth]{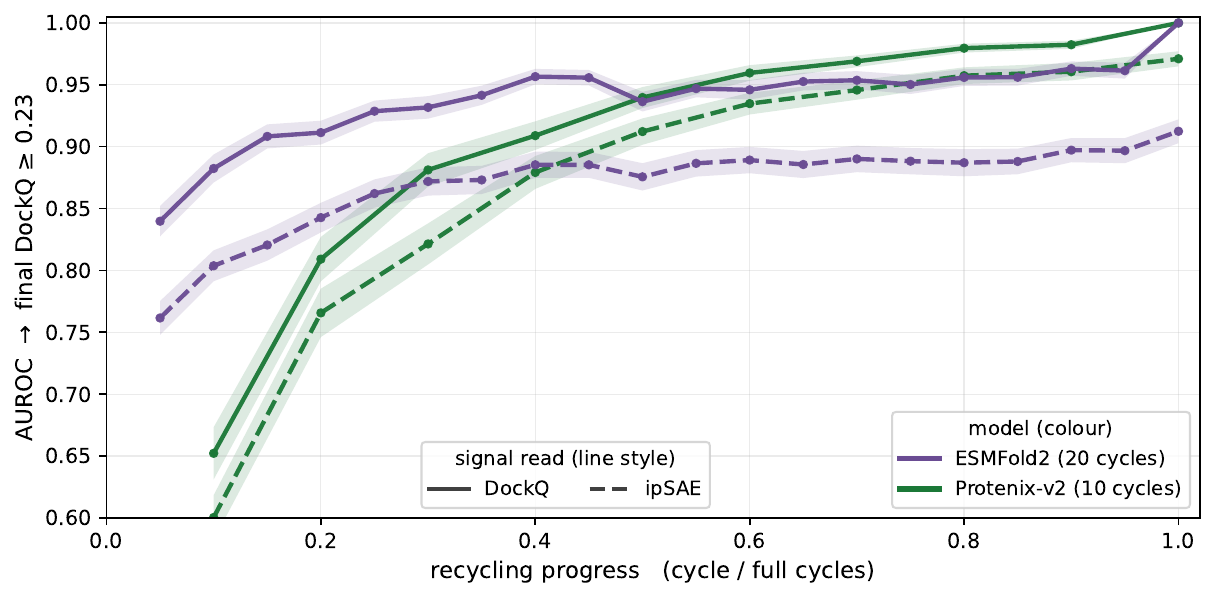}
    \caption{intermediate AG$\rightarrow$AB ipSAE and DockQ during recycling as predictive of final DockQ}
    \label{fig:cycle-auroc-vs-final-dockq}
\end{figure}

\Cref{fig:cycle-auroc-vs-final-dockq} shows that trunk outputs carry information about the final DockQ score even in early cycles. We ran diffusion and confidence head once after every recycle and looked at the predictiveness for the final structure being acceptable (DockQ $>$ 0.23).

\begin{figure}[t]
    \centering
    \begin{subfigure}[t]{0.75\linewidth}
        \centering
        \includegraphics[width=\linewidth]{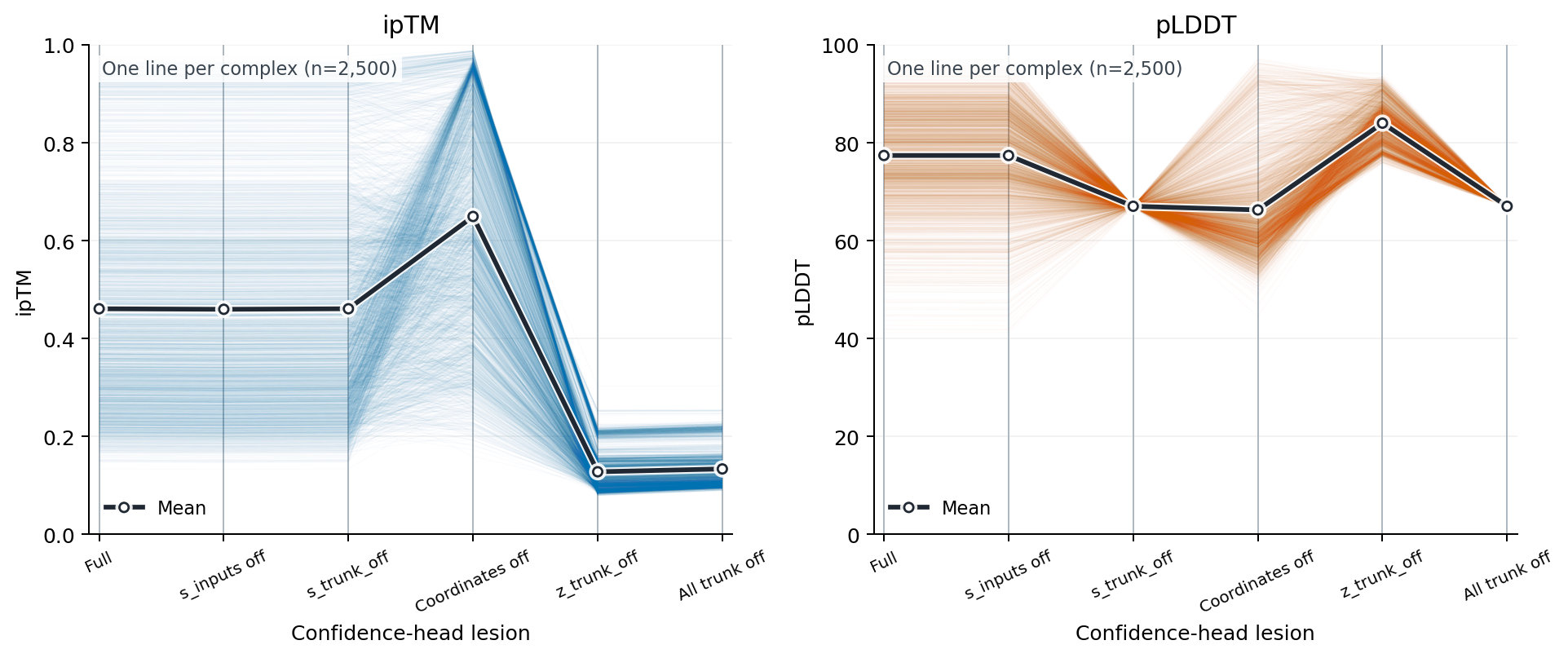}
        \caption{Protenix-v2 confidence head input ablations}
        \label{fig:px-coordinate-perturbations}
    \end{subfigure}
    \qquad
    \begin{subfigure}[t]{0.75\linewidth}
        \centering
        \includegraphics[width=\linewidth]{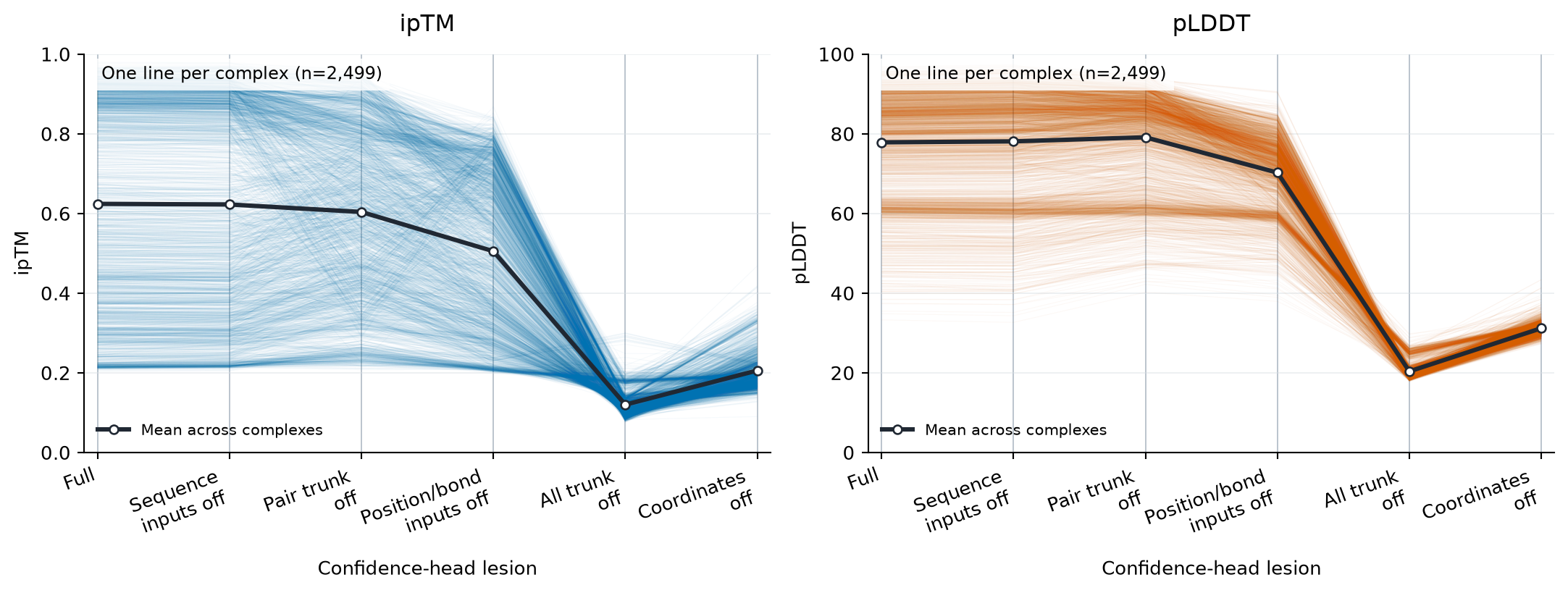}
        \caption{ESMFold2 confidence head input ablations.}
        \label{fig:esmfold2-confidence-lesions}
    \end{subfigure}
    \caption{\textbf{Ablations of inputs to Protenix-v2 and ESMFold2 confidence heads.} The 
    confidence head was re-evaluated on 32 sampled structures for each of 2,500 antibody–antigen complexes randomly selected from SAbDab after selectively ablating input embeddings (\(s_{\mathrm{inputs}}\)), trunk single representations (\(s_{\mathrm{trunk}}\)), trunk pair representations (\(z_{\mathrm{trunk}}\)), or the complete trunk or coordinate pathways. \textbf{Left:} ipTM under each ablation; colored lines show per-complex behavior with points denoting a mean across complexes. \textbf{Right:} pLDDT under each ablation. These results identify the trunk pair representation as the dominant input supporting interface confidence for Protenix-v2.
    }
    \label{fig:confidence-ablations}
\end{figure}

\Cref{fig:px-coordinate-perturbations} shows the confidence metrics of Protenix-v2 and ESMFold2 samples under different input ablations. For Protenix-v2, we observe limited impact for ablation of trunk single representations. While we see an effect on ablating coordinates entirely, when accounting for our observation of the limited effect of noising the coordinates on confidence scores in Figure \ref{fig:coord-perturb},  we hypothesise this is attributable to the ablation being a severe perturbation which is out of distribution for the confidence head. Ablating the trunk pair representation entirely collapses the ipTM. For the pLDDT prediction, the single representation is the primary input, which is unsurprising given the architectural design of confidence heads which primarily operate on this input.

\begin{figure}
    \centering
    \includegraphics[width=\linewidth]{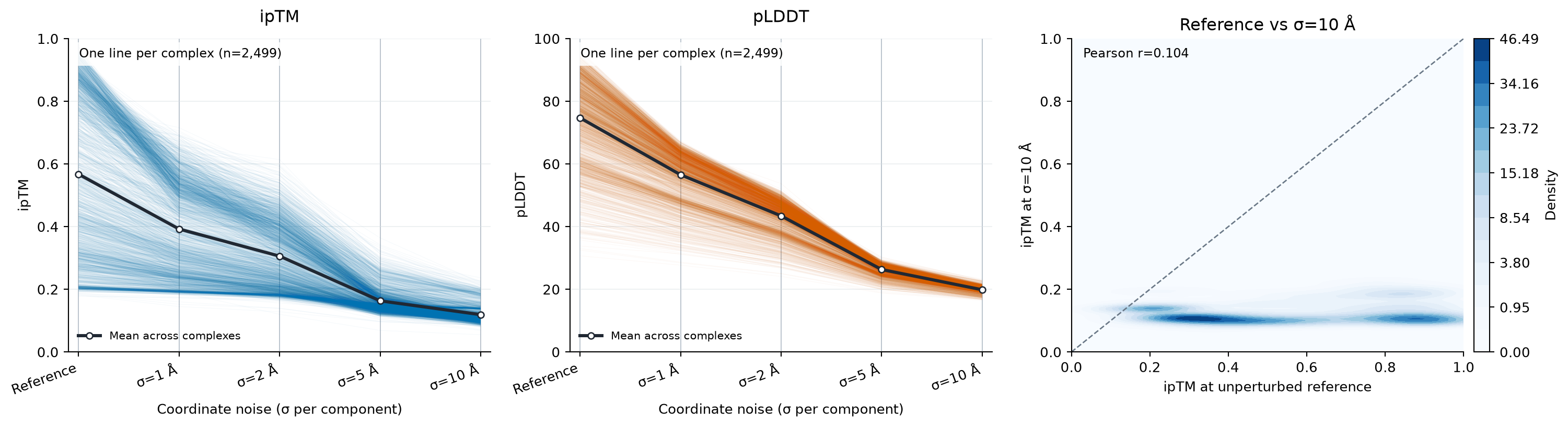}
    \caption{\textbf{Perturbation of structural input to ESMFold2 confidence head.} Trunk representations were fixed while the confidence head scored perturbed structures for 2,499 antibody–antigen complexes randomly sampled from SAbDab. Perturbations comprised independent Gaussian noise with per-coordinate standard deviation \(\sigma \in \{1,2,5,10\}\,\text{Å}\). \textbf{Left:} ipTM across perturbations; traces denote individual complexes and black line denotes mean. \textbf{Centre:} pLDDT across perturbations; traces denote individual complexes and black line denotes mean. \textbf{Right:} Unperturbed versus perturbed ipTM for \(\sigma=10\,\text{Å}\); dashed diagonal indicates perfect agreement.}
    \label{fig:esm-confidence-perturb}
\end{figure}

\Cref{fig:esmfold2-confidence-lesions} shows the same experiment performed on the ESMFold2 confidence head inputs. Here, unlike for Protenix-v2, we observe strong co-ordinate dependence with ipTM prediction collapsing for co-ordinate ablation. Similarly, in Figure \ref{fig:esm-confidence-perturb} we observe monotonic degradation in ipTM and pLDDT confidence scores with increased noising of the structural input. However, we note ESMFold2 trains the confidence head with 20\% dropout on the pair representation to deliberately reduce the dependence on the pair \citep{biohub2026languagemodeling} so this difference is unsurprising.

\section{Training the Confidence CycleCritic}
\label{sec:app-training}

The host folding model is \emph{frozen} throughout. For each training
structure we run the model once through its complete inference schedule and
record two things: the pair representation
$z_c \in \mathbb{R}^{N \times N \times d_\text{pair}}$ at \emph{every} recycle
$c$, and the \emph{final} structural metric obtained after the full schedule
completes --- ipSAE, ipTM, pTM or pLDDT computed on the output of the confidence head.
The metric after all recycling loops is the target; the per-cycle representations are the inputs.

The CC consumes $z_c\in \mathbb{R}^{N \times N \times d_\text{pair}}$ at recycle $c$ together with learned chain-identity and recycle embeddings and emits a categorical distribution $P_c \in \Delta^K$ over $K$ bins
tiling $[0,1]$, with scalar prediction $s_c = \sum_k P_{c,k}\, b_k$ for bin
centres $b_k$. Three terms are combined, each evaluated independently at every
recycle and then averaged:
\begin{equation}
  \mathcal{L} \;=\; \lambda_\text{nll}\,\mathcal{L}_\text{NLL}
              \;+\; \lambda_\text{rank}\,\mathcal{L}_\text{rank}
              \;+\; \lambda_\text{auroc}\,\mathcal{L}_\text{AUROC}.
  \label{eq:cc-loss-train}
\end{equation}
$\mathcal{L}_\text{NLL}$ is the cross-entropy against the bin containing the
target metric. $\mathcal{L}_\text{rank}$ is a RankNet \citep{burges2005learning} term over a mini-batch of
$B$ structures,
\begin{equation}
  \mathcal{L}_\text{rank}
    \;=\; \frac{1}{|\mathcal{P}|} \sum_{(i,j) \in \mathcal{P}}
          \operatorname{softplus}\!\left( -\frac{s_i - s_j}{\tau} \right),
  \qquad
  \mathcal{P} = \{(i,j) : y_i > y_j\},
  \label{eq:cc-rank-train}
\end{equation}
a smooth surrogate for Spearman correlation over the strictly ordered pairs
$\mathcal{P}$ and $\tau=0.1$. $\mathcal{L}_\text{AUROC}$ is the Wilcoxon--Mann--Whitney
surrogate \citep{burges2005learning, yan2003wmw} at a quality bar $\theta=0.5$, applied to the exceedance mass
$\sum_{b_k > \theta} P_{i,k}$ rather than to $s_i$, so that it optimises the
separability the gate actually thresholds on.

Applying the pairwise terms per recycle trains the head to rank candidates
correctly at every point along the trajectory, not only once recycling has
converged.

We train for up to 30 epochs of 100 optimiser steps, each coupling $B$ structures,
using AdamW ($\beta = (0.9, 0.98)$, weight decay $10^{-3}$) with cosine
annealing and gradient clipping at 1.0. Checkpoints are selected on the
mean-over-recycles validation Spearman correlation.

The hyperparameters $\lambda_{\text{rank}}$ and $\lambda_{\text{AUROC}}$ are found via hyperparameter optimization with the search space as defined in \Cref{tab:cc-search-space}.
The objective of the sweep was mean spearman correlation of the CC prediction to the target metric.

\begin{table}[t]
\centering
\caption{\textbf{CycleCritic hyperparameter search space.} Both substrates were searched
with Bayesian optimisation under Hyperband early termination
($\text{min\_iter}=8$, $\eta=3$), maximising the mean Spearman correlation across
recycles on a held-out validation set. The Protenix column applies identically to
the gating sweep and all four design sweeps; the ESMFold2 column describes the
experimental-fast design sweep.}
\label{tab:cc-search-space}
\footnotesize
\setlength{\tabcolsep}{5pt}
\begin{tabular}{|l|c|c|}
\hline
Hyperparameter & Protenix-v2 (stock) & ESMFold2 exp.-fast \\
\hline\hline
\multicolumn{3}{|l|}{\textbf{Searched}}\\
\hline
Architecture & \multicolumn{2}{c|}{\texttt{ffn}, \texttt{swiglu}} \\
\hline
$d_\text{hid}$ & 16, 32, 64, 128, 256 & 64, 128, 256 \\
\hline
$d_\text{lat}$ & 16, 32, 64, 128, 256, 384 & 64, 128, 256 \\
\hline
$n_\text{hid}$ & \multicolumn{2}{c|}{1, 2, 3} \\
\hline
Bins $K$ & 16, 32, 64 & 16, 32, 64, 128 \\
\hline
$p_\text{drop}$ & 0.0, 0.1 & 0.0, 0.1, 0.2 \\
\hline
Learning rate & log-$\mathcal{U}[4\!\times\!10^{-4}, 6\!\times\!10^{-3}]$ & log-$\mathcal{U}[10^{-4}, 3\!\times\!10^{-3}]$ \\
\hline
$\lambda_\text{NLL}$ & 1 & 1 \\
\hline
$\lambda_\text{rank}$ & \multicolumn{2}{c|}{log-$\mathcal{U}[0.1, 12]$} \\
\hline
$\lambda_\text{auroc}$ & 0, 0.3, 1 & 0, 0.3, 1, 3 \\
\hline
\texttt{rank\_loss} & \multicolumn{2}{c|}{true, false} \\
\hline
\texttt{weighted\_pool} & \multicolumn{2}{c|}{true, false} \\
\hline
\texttt{nll\_over\_samples} & \multicolumn{2}{c|}{true, false} \\
\hline
\texttt{head\_on\_pool} &  false & true, false \\
\hline\hline
\multicolumn{3}{|l|}{\textbf{Fixed}}\\
\hline
Optimiser & \multicolumn{2}{c|}{AdamW, $\beta=(0.9,0.98)$, weight decay $10^{-3}$} \\
\hline
Schedule & \multicolumn{2}{c|}{cosine annealing over all steps, gradient clip 1.0} \\
\hline
Epochs $\times$ steps & \multicolumn{2}{c|}{$30 \times 100$} \\
\hline
$\lambda_\text{nll}$, $\tau$, $\theta$ & \multicolumn{2}{c|}{$1.0$, $0.1$, $0.5$} \\
\hline
Rank batch $B$ & 4 & 12 \\
\hline
Run budget & 300 (gating), 200 (design) & 800 \\
\hline
\end{tabular}
\end{table}

\begin{table}[t]
\centering
\caption{\textbf{CycleCritic head configurations, grouped by regime and substrate.}
Gating heads score native sequences over the host model's full recycling
schedule; design heads score CDR-noised sequences at four cycles.
$d_\text{pair}$ is the trunk pair-representation width, $d_\text{asym}$ the
asymmetric chain-pair embedding, and \emph{w-pool} / \emph{h-pool} the
\texttt{weighted\_pool} and \texttt{head\_on\_pool} flags.}
\label{tab:cc-heads}
\footnotesize
\setlength{\tabcolsep}{3.5pt}
\begin{tabular}{|l|l|rr|rrrr|rr|cc|r|}
\hline
Target & Arch. & $d_\text{pair}$ & Cyc. & $d_\text{hid}$ & $d_\text{lat}$ &
$n_\text{hid}$ & Bins & $d_\text{asym}$ & $p_\text{drop}$ &
w-pool & h-pool & Params \\
\hline\hline
\multicolumn{13}{|l|}{\textbf{Gating}}\\
\hline
\multicolumn{13}{|l|}{\quad\textit{Protenix-v2 (stock trunk), 10 cycles}}\\
\hline
ipSAE & \texttt{ffn} & 256 & 10 & 256 & -- & 3 & 16 & 4 & 0.1 & \cmark & \xmark & 205{,}457 \\
\hline
\hline\hline
\multicolumn{13}{|l|}{\textbf{Design}}\\
\hline
\multicolumn{13}{|l|}{\quad\textit{Protenix-v2 (stock trunk), 4 cycles}}\\
\hline
ipSAE & \texttt{swiglu} & 256 & 4 & 32 & 128 & 1 & 32 & 4 & 0.1 & \cmark & \xmark & 74{,}793 \\
\hline
ipTM & \texttt{swiglu} & 256 & 4 & 64 & 384 & 2 & 32 & 4 & 0.1 & \cmark & \xmark & 238{,}473 \\
\hline
pTM & \texttt{swiglu} & 256 & 4 & 16 & 128 & 1 & 32 & 4 & 0.1 & \cmark & \xmark & 72{,}217 \\
\hline
pLDDT & \texttt{swiglu} & 256 & 4 & 16 & 384 & 1 & 16 & 4 & 0.1 & \cmark & \xmark & 214{,}025 \\
\hline
\multicolumn{13}{|l|}{\quad\textit{ESMFold2 experimental-fast-cutoff2021/2025, 4 cycles}}\\
\hline
ipTM & \texttt{ffn} & 256 & 4 & 256 & -- & 1 & 16 & 4 & 0.2 & \cmark & \cmark & 73{,}849 \\
\hline
ipSAE & \texttt{ffn} & 256 & 4 & 128 & -- & 1 & 32 & 4 & 0.1 & \cmark & \cmark & 39{,}305 \\
\hline
\end{tabular}
\end{table}

\begin{table}[t]
\centering\small
\begin{tabular}{|l|r|r|r|r|}
\hline
Target & lr & $\lambda_\text{rank}$ & $\lambda_\text{auroc}$ & $b_\text{rank}$ \\
\hline\hline
\multicolumn{5}{|l|}{\textbf{Gating}}\\
\hline
\multicolumn{5}{|l|}{\quad\textit{Protenix-v2}}\\
\hline
ipSAE & $3.46\times10^{-3}$ & 0 & 0 & - \\
\hline\hline
\multicolumn{5}{|l|}{\textbf{Design}}\\
\hline
\multicolumn{5}{|l|}{\quad\textit{Protenix-v2}}\\
\hline
ipSAE & $1.67\times10^{-3}$ & 1.431 & 1 & 4 \\
\hline
ipTM & $3.36\times10^{-3}$ & 4.665 & 1 & 4 \\
\hline
pTM & $5.62\times10^{-3}$ & 4.818 & 0 & 4 \\
\hline
pLDDT & $5.20\times10^{-3}$ & 2.387 & 1 & 4 \\
\hline
\multicolumn{5}{|l|}{\quad\textit{ESMFold2 exp.-fast-cutoff2021/2025}}\\
\hline
ipTM & $2.14\times10^{-3}$ & 6.839 & 1 & 12 \\
\hline
ipSAE & $2.19\times10^{-3}$ & 11.562 & 0.3 & 12 \\
\hline
\end{tabular}
\caption{CycleCritic training hyperparameters for deployed models the optimal parameters were also found via Bayesian optimization.}
\label{tab:cc-head-hyperparams}
\end{table}

\subsection{Architecture details for the Confidence CycleCritic}

For each trunk pair representation $z_{ij}\in\mathbb{R}^{d_{\mathrm{pair}}}$, learned
row-chain, column-chain, and recycling-cycle embeddings are concatenated and
LayerNorm is applied:
\begin{equation}
  x_{ij} = \operatorname{LayerNorm}\!\left(
    \left[z_{ij}; e_i^{\mathrm{chain}}; e_j^{\mathrm{chain}};
    e^{\mathrm{cycle}}\right]
  \right).
\end{equation}
Thus, $x_{ij}\in\mathbb{R}^{d_{\mathrm{in}}}$, where
$d_{\mathrm{in}}=d_{\mathrm{pair}}+3d_{\mathrm{asym}}$. The same FFN or SwiGLU operation
$E_{\mathrm{pair}}$ is applied independently to every $x_{ij}$, producing one
latent representation per pair. The pair axis is reduced by a
permutation-invariant mean, optionally weighted by a learned confidence, and
the output head is applied either before or after pooling. The head maps its
input to $K$ logits with a final linear layer and applies softmax, producing a
categorical distribution $P$ over $K$ equal-width bins of the target in
$[0,1]$. The scalar prediction is the expected bin center,
$\hat{y}=\sum_{k=1}^{K}P_kb_k$, where $b_k$ is the center of bin $k$.

\paragraph{Per-pair activation.}
The FFN uses $n_{\mathrm{hid}}$ linear--GELU layers. Its final per-pair latent
has width $d_{\mathrm{hid}}$:
\begin{equation}
  h_{ij}^{(0)}=x_{ij},\qquad
  h_{ij}^{(r)}=\operatorname{GELU}\!\left(
    W_rh_{ij}^{(r-1)}+b_r
  \right),\quad r=1,\ldots,n_{\mathrm{hid}}.
\end{equation}
Thus,
$E_{\mathrm{pair}}(x_{ij})=h_{ij}^{(n_{\mathrm{hid}})}
\in\mathbb{R}^{d_{\mathrm{hid}}}$. Its output head is
\begin{equation}
  \operatorname{Head}_{\mathrm{FFN}}(h)
  =\operatorname{softmax}(W_{\mathrm{out}}h+b_{\mathrm{out}}),
  \qquad
  W_{\mathrm{out}}\in\mathbb{R}^{K\times d_{\mathrm{hid}}}.
\end{equation}

The SwiGLU encoder produces a per-pair latent of width $d_{\mathrm{lat}}$
through a gated projection:
\begin{equation}
  E_{\mathrm{pair}}(x_{ij})
  =(W_{\mathrm{up}}x_{ij}+b_{\mathrm{up}})
  \odot\operatorname{SiLU}(W_{\mathrm{gate}}x_{ij}+b_{\mathrm{gate}}),
  \qquad
  W_{\mathrm{up}},W_{\mathrm{gate}}
  \in\mathbb{R}^{d_{\mathrm{lat}}\times d_{\mathrm{in}}}.
\end{equation}
Its output head applies an $n_{\mathrm{hid}}$-layer SiLU MLP of width
$d_{\mathrm{hid}}$, followed by a linear map to $K$ logits:
\begin{equation}
  \operatorname{Head}_{\mathrm{SwiGLU}}(h)
  =\operatorname{softmax}\!\left(
    W_{\mathrm{out}}
    \operatorname{MLP}_{\mathrm{SiLU}}^{(n_{\mathrm{hid}})}(h)
    +b_{\mathrm{out}}
  \right),
  \qquad
  W_{\mathrm{out}}\in\mathbb{R}^{K\times d_{\mathrm{hid}}}.
\end{equation}
The MLP maps $\mathbb{R}^{d_{\mathrm{lat}}}$ to
$\mathbb{R}^{d_{\mathrm{hid}}}$; its first weight matrix has shape
$d_{\mathrm{hid}}\times d_{\mathrm{lat}}$, and subsequent matrices have
shape $d_{\mathrm{hid}}\times d_{\mathrm{hid}}$. Thus, both architectures
end in a linear map to $K$ logits, while the SwiGLU head includes the
preceding SiLU MLP.

\paragraph{Pooling operation.}
The pooling operation is either a learned weighted mean or a uniform mean.
With weighted pooling, a linear confidence head predicts $c_{ij}$ from each
pair latent and assigns weight $w_{ij}=\sigma(c_{ij})$; without weighting,
$w_{ij}=1$. These weights are used in either aggregation order below.

\paragraph{Head placement.}
\texttt{head\_on\_pool} determines the ordering of the head and pooling
operations. When it is false, the head is applied to each pair before
pooling; when it is true, the pair representations are pooled before applying
the head:
\begin{itemize}
  \item \texttt{False}:
    $\operatorname{Pool}\!\left(
      \operatorname{Head}\!\left(E_{\mathrm{pair}}(x_{ij})\right)
    \right)$.
  \item \texttt{True}:
    $\operatorname{Head}\!\left(
      \operatorname{Pool}\!\left(E_{\mathrm{pair}}(x_{ij})\right)
    \right)$.
\end{itemize}

\section{Dataset construction}
\label{app:datasets}

All datasets were generated by running a folding model over
antibody--antigen complexes from SAbDab and recording the per-recycle pair representation
$z_{\text{cycles}} \in \mathbb{R}^{C \times N \times N \times D}$,
together with \texttt{asym\_id} (per-token chain index), \texttt{is\_ab}
(antibody mask), and decoded confidence labels. 
Each complex is decoded
$N_{\text{decode}}$ times under independent diffusion noise; reported labels are
the mean over decodes, with the per-decode values retained.

Two regimes are used. \emph{Gating} datasets fold native sequences and are used
to predict refold quality from the trunk state. \emph{Design} datasets fold
CDR-corrupted sequences and are used to guide hallucination-based design
trajectories.

For splitting, we use the epitope similarity proposed in \cite{isomorphic2026isodde}. We use mmseqs clustering at 40 percent  similarity. The resulting clusters are split into validation and training with a 1:7 ratio.

\subsection{Gating datasets}
\label{app:gating}

The parameters for generating the gating dataset for Protenix-v2 are summarized in \Cref{tab:gate-data}.

\noindent
CycleCritic used the Protenix-v2  code and weights from its \href{https://github.com/bytedance/Protenix}{github repo} as available under the Apache 2.0 license downloaded on date April 8th 2026.
\begin{table}[t]
\centering
\caption{\textbf{Parameters used to generate the Protenix-v2 gating dataset.}
Each native complex is folded once using the full recycling schedule, with
pair representations captured after every recycle. Antigen and
antibody-framework MSAs are provided.}
\label{tab:gate-data}
\footnotesize
\setlength{\tabcolsep}{6pt}
\begin{tabular}{@{}lc@{}}
\toprule
Parameter & Value \\
\midrule
Host model              & Protenix-v2 \\
MSA                      & Yes \\
Recycles, $C$            & 10 \\
$N_{\mathrm{step}}$      & 200 \\
$N_{\mathrm{decode}}$    & 5 \\
Token cap (train / val)  & 600 / 1{,}000 \\
Complexes (train / val)  & 3{,}498 / 500 \\
\bottomrule
\end{tabular}
\end{table}

\subsection{Design datasets}
\label{app:design}

For training the cCC for design, we explicitly also noise the CDRs in order for the CC to not be out of distribution at the start of design.
We observed that this had a great positive effect on generation.

\paragraph{CDR noise model}
A soft position-specific scoring matrix replaces the one-hot residue encoding (\texttt{restype}) at
CDR positions. For CDR position $i$ with native one-hot
$e_i \in \mathbb{R}^{20}$,
\begin{align}
\mathbf{v}_i &= 10\, e_i + \sigma\,\varepsilon_i,
  \qquad \varepsilon_i \sim \mathcal{N}(0, I_{20}) \\
\text{PSSM}_i &= \operatorname{softmax}(\mathbf{v}_i).
\end{align}
Framework and antigen positions retain their one-hot encoding. For all
$\sigma > 0$ the cysteine logit is set to $-10^{6}$ prior to
the softmax. Each complex contributes one $\sigma = 0$ anchor and
$k_{\text{noised}} = 4$ noised draws, giving five items per complex.

Inspired by \citet{karras2022elucidating}, the noise scale is drawn from a clipped log-normal,
$\sigma = \min(\exp(u),\,\sigma_{\max})$ with
$u \sim \mathcal{N}(P_{\text{mean}},\,P_{\text{std}})$.

\subsubsection{Protenix/ Mosaic}

\begin{table}[ht]
\centering
\begin{tabular}{ll}
\toprule
Parameter & Value \\
\midrule
Model & Protenix-v2 \\
Recycles $C$ & 4 \\
Diffusion steps $N_{\text{step}}$ & 200 \\
Decodes $N_{\text{decode}}$ & 5 \\
$(P_{\text{mean}}, P_{\text{std}}, \sigma_{\max})$ & $(1.15,\ 0.45,\ 12.0)$ \\
Median $\sigma$ & $\approx$3.2 \\
Cropping & epitope-centered, pad 30 \\
Max tokens & 600 \\
\bottomrule
\end{tabular}
\end{table}

The soft PSSM is written into the \texttt{restype} feature. 
The \texttt{profile} feature receives
the rank-one query-row correction
$\Delta\text{profile}_i = (\text{PSSM}_i - \text{old}_i)/N_{\text{msa}}$; the
\texttt{msa} feature is unmodified. Columns $20{+}$ of \texttt{restype} are
zeroed.

Schedule constants were calibrated on 50 complexes (23 SAbDab, 27 VHH) whose
$\sigma = 0$ anchors reached ipTM $\geq 0.70$. Both regimes are flat to
$\sigma = 2$; VHH then falls to $0.65$ of anchor by $\sigma = 3$, while SAbDab
decays to a plateau near $0.66$.

We sample $u \sim \mathcal{N}(1.15, 0.45^2)$ and set
$\sigma = \min(\exp(u), 12)$.
The resulting median noise scale is approximately $3.2$,
with the central one-standard-deviation interval approximately
$[2.0, 5.0]$.

Each complex contributes five folds: one $\sigma = 0$ anchor and four draws
from the schedule. Training uses 7,089 complexes (35,445 folds) and validation
100 complexes (500 folds).

\subsubsection{ESMFold2}

\begin{table}[ht]
\centering
\begin{tabular}{ll}
\toprule
Parameter & Value \\
\midrule
Model & \texttt{ESMFold2-Experimental-Fast-cutoff2021/2025} \\
\texttt{lm\_dropout} & 0.5 \\
\texttt{num\_loops} & 3 (4 carry cycles) \\
\texttt{num\_sampling\_steps} & 20 \\
Recycles $C$ captured & 4 \\
$(P_{\text{mean}}, P_{\text{std}}, \sigma_{\max})$ & $(0.9,\ 0.7,\ 12.0)$ \\
Median $\sigma$ & $\approx 2.5$ \\
Cropping & epitope-centered, pad 30 \\
Max tokens & 600 \\
\bottomrule
\end{tabular}
\end{table}

\textbf{Binder construction.} The binder is a single suffix chain,
$\text{VH} + \texttt{GGGSGGGSGGGSGGGS} + \text{VL}$, or VH alone for
VHH/nanobody. Complexes are folded as \texttt{antigen | binder}.

\textbf{Filtering.} Retained complexes have a heavy chain, exactly one antigen
chain, and a defined epitope. The antigen is cropped to
$[\min(\text{epitope}) - 30,\ \max(\text{epitope}) + 30]$, and complexes whose
cropped target $+$ binder exceeds 600 tokens are dropped.

Calibration on this model gives ipTM collapse across $\sigma \in [1,4]$, from
$0.8$--$0.9$ at $\sigma = 0$ to $\approx 0.2$ for $\sigma \geq 4$.

\textbf{Splits.} Train $=$ SAbDab train $+$ VHH train; val $=$ SAbDab val $+$
VHH val; reusing the same epitope-cluster splits. Val
 are 200 complex-folds drawn with \texttt{WeightedSampler}
(\texttt{weights\_epitope\_then\_cdr}, without replacement, )
Training
uses all valid complexes.

\section{Structure CycleCritic}
\label{app:struct-cc}
The structure cycle critic (sCC) scores a candidate at recycle $c$ from two quantities that are already available before diffusion is run: the trunk's current belief about pairwise geometry, and the designed backbone the refold is supposed to reproduce. Both are expressed on the same distance basis, which is what allows them to be compared pair by pair.

\subsection{Analytic baseline}
\label{app:scc_baseline}
As a zero-parameter baseline we use the analytic distogram criterion (Algorithm~\ref{alg:nll-iface}), which based on the predicted distogram and the designed backbone score a refold on a simple statistic: the mean negative log-likelihood, under $p_{ij}$, of the bin containing the reference distance, taken over cross-role pairs closer than 12~\AA. A candidate is discarded when this statistic exceeds a per-recycle threshold, subject to a minimum-recycle guard that suppresses kills before the pose has had a chance to form.

\begin{algorithm}[t]
  \caption{Interface distogram NLL at recycle $c$}
  \label{alg:nll-iface}
  \begin{algorithmic}[1]
  \Require $z^{(c)} \in \mathbb{R}^{N \times N \times d}$, trunk pair
    representation at recycle $c$
  \Require $D \in \mathbb{R}^{N \times N}$, pairwise distances between the
    representative atoms of the \emph{design} backbone
  \Require chain roles $\mathrm{r}(i) \in \{\text{binder}, \text{target}\}$;
    interface cutoff $\tau$ (12\,\AA); bin boundaries $e_1 < \dots < e_{B-1}$
    of the trunk's distogram head
  \Ensure $\mathrm{nll}_{\text{iface}}$, the mean negative log-likelihood of the
    design's interface distances under the trunk's own belief
    
  \Statex
  
  \State $\ell \gets \mathrm{DistogramHead}(z^{(c)})$
  \Comment{The distogram head, exactly as the trunk runs it, frozen;}
  \State $\log p \gets \operatorname{log\text{-}softmax}(\ell)$
    \Comment{symmetrised internally over the bin axis.}

  \Statex
  
  \State $b_{ij} \gets \sum_{k=1}^{B-1} \mathbf{1}[D_{ij} > e_k]$
     \Comment{Bin the observed backbone distances.}
  \Statex
  \Statex
  \Comment{Interface pairs: cross-role, in contact, both tokens resolved.}
  \State $M \gets \{(i,j) : \mathrm{r}(i) \neq \mathrm{r}(j)
    \;\wedge\; i \neq j \;\wedge\; D_{ij} < \tau \;\wedge\; i,j \text{ resolved}\}$
  \If{$M = \varnothing$} \State \Return $\varnothing$
    \Comment{no interface: the gate falls back or abstains} \EndIf
 
  \Statex
 
  \State $\mathrm{nll}_{\text{iface}} \gets
    -\dfrac{1}{|M|} \displaystyle\sum_{(i,j) \in M} \log p_{ij}\big[b_{ij}\big]$
     \Comment{Average the surprise over the interface.}
  \State \Return $\mathrm{nll}_{\text{iface}}$
  \end{algorithmic}
  \end{algorithm}

\subsection{Learned sCC: Architecture and training}
\label{app:learned_scc}
In Fig.~\ref{fig:sccc-architecture} the architecture of the learned sCC is illustrated. A more detailed description is outlined below.

\paragraph{Inputs.} In addition to the predicted distogram $p_{ij}$, learned sCC from the designed backbone takes the C$_\beta$ distance $d_{ij}$ of the same token pair (C$_\beta$ constructed from N, C$_\alpha$ and C, so the reference is backbone-only and independent of side chains) and encodes it on \emph{those same bins} as a soft code $q_{ij}$. Two flags accompany each pair: whether both tokens are resolved in the reference, and whether $d_{ij}$ fell in the open-ended far bin, where the head cannot distinguish 22~\AA{} from 80~\AA.
\paragraph{Pair selection.} Scoring all $N^2$ pairs is unnecessary and, for a pooled statistic, actively harmful: interface contacts are well under $1\%$ of the grid, so a uniform draw would contain almost none of them. We therefore keep a stratified budget of $10{,}240$ pairs in three bands. Every cross-role pair within 18~\AA{} is retained, so the \emph{interface} band ($<12$~\AA) and the \emph{halo} band (12--18~\AA) are stored complete rather than sampled; the remaining budget is drawn uniformly from the rest of the grid, so the sampled \emph{background} band's mean is an unbiased estimate of the mean over the remaining pairs.
\paragraph{Per-pair encoding.} Each retained pair is described by the concatenation of $p_{ij}$, $q_{ij}$, the two flags, a learned embedding of the role pair (binder--binder, binder--target, target--target) and a learned embedding of the recycle index, together with the scalar agreement $a_{ij} \;=\; -\log \left(p_{ij}\cdot q_{ij}\right)$, the surprise of the reference distance under the trunk's own prediction. Averaging $a_{ij}$ over the interface only would reproduce the analytic baseline, the distogram NLL. This gives a 139-dimensional token, layer-normalized, which a single SwiGLU shared across all pairs maps to a 128-dimensional latent.
\paragraph{Pooling and readout.} Latents are averaged within each band, each normalized by its own occupancy so that a band contributes equally regardless of how many pairs it holds. Alongside them we append the mean, maximum and 95th percentile of $a_{ij}$ within each band, computed directly as $a_{ij}$ above and bypassing the SwiGLU, so that a purely analytic summary of interface agreement remains a linear readout of the pooled vector. The concatenation (393 dimensions) passes through a 128-unit layer with SiLU activation and dropout $0.1$, and a final linear layer with a softmax over 40 bins uniformly covering 0--20~\AA{} yields a predicted distribution over sc-RMSD. Gating uses the tail mass $P(\text{sc-RMSD}<3~\text{\AA})$, the sum of the first six bins; 3~\AA{} is chosen to fall exactly on a bin edge. The head has $96{,}946$ parameters, negligible against a single trunk recycle.
\paragraph{Training.} For each item we record the sc-RMSD of five diffusion decodes at full depth. At every recycle the sCC predicts the distribution of that same full-depth outcome. The loss is the negative log-likelihood of those five outcomes under the prediction, averaged over decodes and summed over recycles with weights decaying linearly from $3$ at recycle $0$ to $1$ at recycle $9$, which concentrates capacity on the early recycles where a rejection is worth making.

\input{sccc_architecture}

\subsection{Dataset}
\label{app:vhh_dataset}

We train and evaluate the learned sCC on a curated set of VHH complexes derived from the Protein Data Bank. The corpus comprises 1444 distinct PDB entries, decomposed into 2635 VHH-antigen structures. Train/validation partitioning is epitope-cluster-based. The split targets $90/10$ at the cluster level: 574 training and 64 validation clusters, with no cluster shared between sides. This yielded 2196 training and 439 validation VHH-antigen pairs, or 16.7\% by structure count rather than 10\% due to the validation clusters being arbitrarily larger than those in the training set. A maximum sequence threshold of 1000 tokens per complex was enforced, yielding 1994 unique structures and 282 held-out validation structures.

\section{Gating binder designs}
\label{app:gating_campaign}

\subsection{Prediction fidelity and analytic CycleCritic}
\label{app:analytic_baseline}
We evaluate the CycleCritic on VHH designs generated as described in \Cref{sec:Gating}.

\begin{figure}
    \centering
    \includegraphics[width=0.9\linewidth]{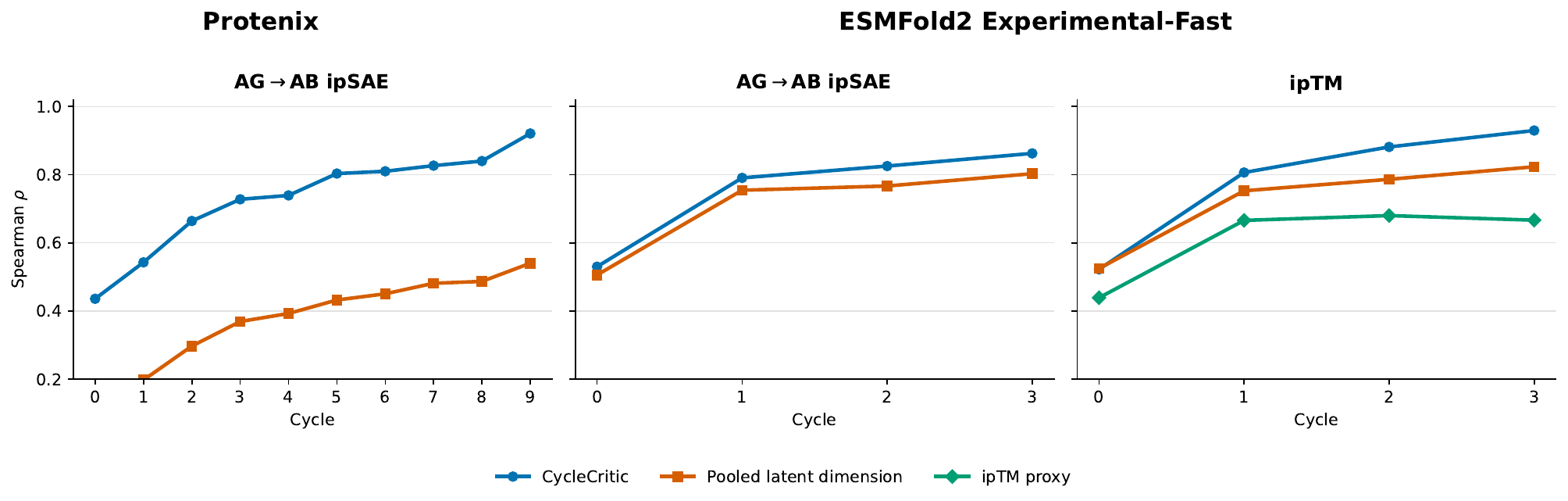}
    \caption{\textbf{Confidence CycleCritic predictions strongly correlate with confidence metrics even at early recycle.} This holds both for Protenix-v2 and ESMFold2. Pooled latent is the baseline described in Figure~\ref{fig:latent-dim-predictiveness}. The baseline ipTM proxy was proposed in \cite{biohub2026languagemodeling}.}
    \label{fig:pred-cycle-esmxfast}
\end{figure}

We first compare the Confidence CycleCritics to the following baselines: 1) The latent dimension in the pair representation that best predicts ipTM or ipSAE and  2) the ipTM proxy as proposed in \citet{biohub2026languagemodeling} (Algorithm 15) for ipTM.
In \Cref{fig:pred-cycle-esmxfast}, we evaluate the Protenix-based gating CC and the ESM-based design model. 
We can see that already early on in recycling the CC is more predictive of the final metric, where gains are most valuable because they allow us to directly apply hallucination or early gating.

\begin{figure}[ht]
    \centering
    \includegraphics[width=.8\linewidth]{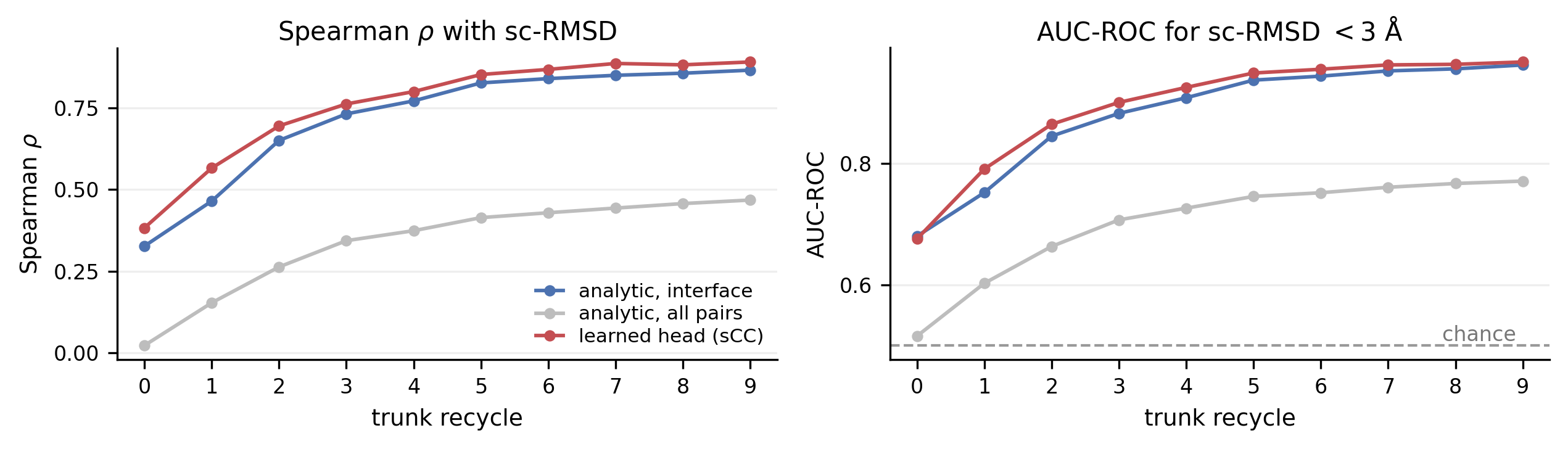}
    \caption{
      \textbf{Structure CycleCritic (sCC) discriminates self-consistency from the 
      second recycle onward.} \textbf{Left:} Spearman
  $\rho$ between the cycle-$c$ sCC scores and the pose's sc-RMSD \textbf{Right:} AUC-ROC of sCC scores
  for the binary call sc-RMSD $<3$\,\AA{}.
    } 
    \label{fig:nll-per-cycle}
  \end{figure}

For the Structure CycleCritic (sCC), we include two analytic critics as baselines: a variant that aggregates likelihood over all inter-chain distances and not only the interface and the one that aggregates only over the interface. In Fig.~\ref{fig:nll-per-cycle} we show correlation of the sCC scores with the self-consistency RMSD of the binder as well as its predictive power (AUROC) per cycle. 
The learned sCC exceeds the analytic baseline at every recycle in both rank agreement and AUROC. The operating point is where the two separate: sweeping a per-cycle schedule to a matched recall of $0.96$, the learned critic saves $0.284$ of the recycling-plus-diffusion budget against $0.209$ for the analytic score. Both critics are considerably stronger at late recycles, but the compute saved by a gate accrues early, so the early-cycle behavior dominates deployment value.

\subsection{Refolding filters}

We refold approximately $126{,}000$ designs with Protenix-v2,
using MSAs for the antigen and antibody framework, $10$ recycles,
$10$ sampling steps of the deterministic probability-flow ODE,
$5$ diffusion samples per trunk pass, and ten independent trunk
seeds per design. 
For each trunk seed, we select the diffusion sample with the highest ipTM among the five samples.
The seed passes when this selected sample satisfies both filters.
Two cheaper refolding baselines are evaluated
on the same designs: Protenix-v2 with $4$ recycles and Protenix-mini~\citep{protenixmini}.

A refold seed \emph{passes} when $\mathrm{ipSAE} > 0.5$ and
antigen-superposed self-consistency RMSD $< 3$\,\AA.
Repeat $0$ is reserved for independent method evaluation and
excluded from the reproducibility label. Each design receives
a $k/9$ label, where $k \in \{0,\dots,9\}$ is the number of
passing refolds among the remaining nine seeds. Across the
campaign, $1.0\%$ of designs pass on all nine reference seeds,
while $4.3\%$ pass on at least one.

\subsection{ProtenixScore baseline}

As a baseline for lightweight, diffusion-free scoring we opted for ProtenixScore~\citep{protenixscore} which is an implementation of the AF3Score~\citep{liu2025af3score} approach for Protenix. ProtenixScore can be used to score a given structure without running the full structure prediction of Protenix models. It runs the trunk for all ten recycles but does not proceed to predict the structure using the diffusion Structure Module. Instead, it passes the structure we are scoring, in this case the designed structure as output by BoltzGen~\citep{boltzgen} into the confidence head which produces its scores. The trunk sees the antigen MSA, a framework-realigned binder MSA, and is templated on the input structure. This templating is per-chain, so no inter-chain distance enter the trunk and the pose of the structure is only given to the confidence head. 

ProtenixScore runs the trunk on the design's own coordinates and skips diffusion, producing confidence scores but no structure. Designs are retained at a fixed threshold of $\texttt{ipsae\_max} > 0.9$. The arm is priced as a single pass over the whole cohort: $227.03$ GPU-hours over $126{,}000$ designs, or $6.51$~s each. Against the nine-seed reproducibility label, the $1{,}704$ declarations contain $1{,}022$ true positives at the $\geq5$-of-$9$ threshold with precision $0.600$, recall $0.374$, delivering $4.50$ reproducible designs per GPU-hour.

\subsection{Gating campaign}

\paragraph{Gates.}
Two early-exit gates operate together. If either gate fires, the refold
stops after the current trunk recycle, before diffusion and confidence-head
evaluation:
\begin{itemize}
  \item \textbf{cCC} is a learned head on the trunk pair representation
  that predicts a binned final-depth ipSAE distribution at every recycle.
  It terminates the refold when the predicted probability
  $P(\mathrm{ipSAE} > 0.5)$ falls below a per-cycle threshold.

  \item \textbf{sCC} is a learned head that uses the trunk's distogram
  and the design backbone's pair distances at every recycle to predict
  the distribution of self-consistency RMSD at full recycling depth.
  It terminates the refold when
  $P(\text{sc-RMSD} < 3\,\text{\AA})$ falls below a
  per-cycle threshold.
\end{itemize}
The gates are kill-only: designs that survive both complete the full
refold and are assessed with the downstream ipSAE and sc-RMSD filters.
The per-cycle thresholds for both critics were calibrated on held-out
SAbDab refolds (282 validation complexes from Sec.~\ref{app:vhh_dataset}) to retain all passing refolds in the calibration set (i.e. 100\% recall). Threshold details are shown in Table~\ref{tab:thresholds}.

\begin{table}[t]
\centering
\small
\begin{tabular}{lcccccccccc}
\toprule
& \multicolumn{10}{c}{recycle} \\
\cmidrule(l){2-11}
& 0 & 1 & 2 & 3 & 4 & 5 & 6 & 7 & 8 & 9 \\
\midrule
sCC kill    & 0.130 & 0.095 & 0.067 & 0.035 & 0.018 & 0.053 & 0.028 & 0.019 & 0.041 & 0.055 \\
\addlinespace
cCC kill    & 0.297 & 0.271 & 0.239 & 0.130 & 0.102 & 0.044 & 0.055 & 0.034 & 0.098 & 0.368 \\
\bottomrule
\end{tabular}
\caption{Deployed per-recycle gating thresholds. The sCC rejects when its predicted
tail mass $P(\text{sc-RMSD}<3\,\text{\AA})$ falls below the bar. The cCC rejects when
$P(\text{ipSAE}>0.5)$ falls below its kill bar; Both gates use
\texttt{min\_cycles} $=1$, permitting a first rejection at recycle $0$.}
\label{tab:thresholds}
\end{table}

\subsection{Results}

The gate terminates $107{,}478$ designs ($85.6\%$ of the evaluated
campaign) before diffusion; $18{,}057$ reach the sampler. The sCC
accounts for $85{,}182$ terminations at a mean zero-based exit index
of $3.6$, and the cCC for $22{,}296$ at $8.8$. Mean trunk depth,
including designs that survive, falls from $10$ to $6.31$ passes
($36.9\%$ fewer trunk passes).

Of the $5{,}401$ designs passing at least one of nine reference
refolds, $4{,}816$ survive the gate ($89.2\%$ pre-diffusion
retention); $172$ are terminated by cCC and $413$ by sCC.
Under the majority criterion ($\geq 5/9$), $2{,}658$
of $2{,}732$ positives survive ($97.3\%$ retention).

Gating survival is not a positive declaration: after diffusion, $2{,}874$
survivors pass both direct ipSAE and sc-RMSD filters. Against the
$\geq 1/9$ label used for precision and recall in
Table~\ref{tab:gate_methods}, $2{,}769$ declarations are true
positives, giving precision $0.96$ and recall $0.51$.

  \begin{table}[ht]
    \centering
    \small
    \setlength{\tabcolsep}{5pt}
    \begin{tabular}{lrrrrrr}
    \toprule
     & & \multicolumn{2}{c}{\textbf{Against the $\geq\!1/9$ label}} & \multicolumn{3}{c}{\textbf{GPU-h per confirmed passer}} \\
    \cmidrule(lr){3-4}\cmidrule(lr){5-7}
    Method & Accepted & Precision & Recall & $\geq\!1/9$ & $\geq\!5/9$ & $9/9$ \\
    \midrule
    ptx-mini & 1{,}056 & 0.78 $\pm$ 0.03 & 0.15 $\pm$ 0.01 & 0.146 & 0.182 & 0.252 \\
    ptx@4 & 2{,}618 & 0.92 $\pm$ 0.01 & 0.45 $\pm$ 0.02 & 0.074 & 0.092 & 0.157 \\
    ptx@10 & 2{,}925 & 0.95 $\pm$ 0.01 & \textbf{0.52 $\pm$ 0.02} & 0.085 & 0.106 & 0.189 \\
    ptx\_score & 1{,}704 & 0.73 $\pm$ 0.03 & 0.23 $\pm$ 0.02 & 0.183 & 0.222 & 0.321 \\
    cc+sccc gated & 2{,}874 & \textbf{0.96 $\pm$ 0.01} & 0.51 $\pm$ 0.02 & \textbf{0.065} & \textbf{0.080} & \textbf{0.145} \\
    \bottomrule
    \end{tabular}
    \caption{Method comparison on the 126,000-design campaign. \emph{Accepted} is the number
    of designs a method declares a passer; precision and recall score that declaration
    against a design that reproduces on at least one of nine held-out Protenix@10 seeds
    ($n=5{,}401$), with $\pm$ half the width of a 95\% chunk-clustered bootstrap interval
    over designs (50-design chunks resampled within target; designs in a chunk share a
    backbone and task and are not independent). The right-hand block divides the same
    measured cost by the passers confirmed at three progressively stricter labels:
    reproducing on at least one seed, at least five, and all nine. Only the label moves
    across those columns - a method's verdict is its own pose against its own bars and does
    not depend on the label. Best value per column in bold.}
    \label{tab:gate_methods}
    \end{table}

\begin{figure}[ht]
  \centering
  \includegraphics[width=\linewidth]{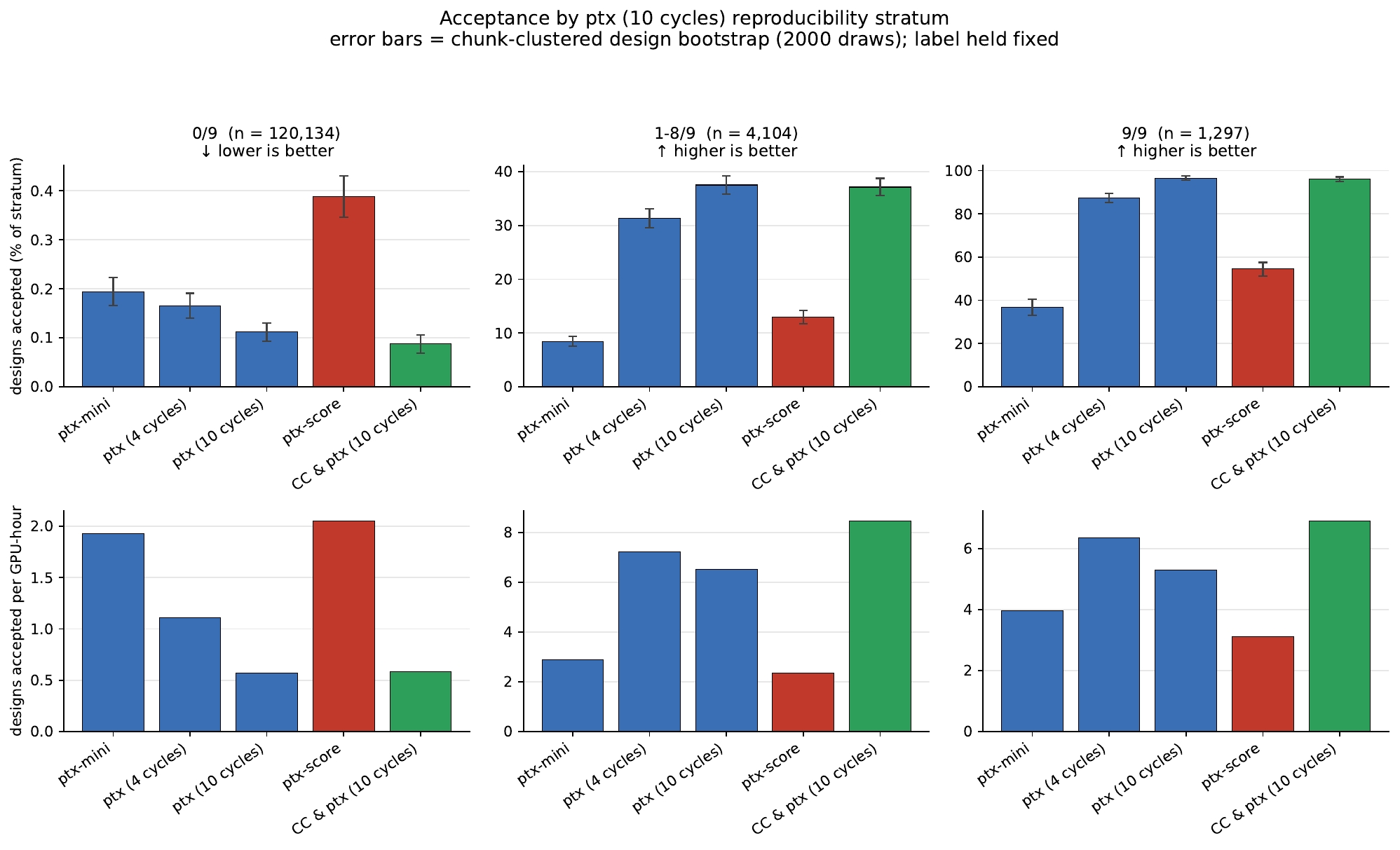}
  \caption{Acceptance by reproducibility stratum. Each design is classified by
  the number of passing refolds among nine held-out seeds (102--110):
  $0/9$ ($n=120{,}134$), $1$--$8/9$ ($n=4{,}104$), or $9/9$
  ($n=1{,}297$). Methods declare independently on seed 101.
  \textbf{Top:} fraction accepted within each stratum.
  \textbf{Bottom:} accepted designs per GPU-hour within each stratum.
  The gate accepts $0.087\%$ of $0/9$ designs while retaining
  $95.99\%$ of $9/9$ designs; ptx@10 accepts $0.112\%$ and
  $96.45\%$, respectively. Error bars are 95\% confidence intervals
  from 2,000 bootstrap draws of 50-design chunks resampled within target.}
  \label{fig:passers_per_gpu}
\end{figure}

\begin{figure}[t]
  \centering
  \includegraphics[width=\textwidth]{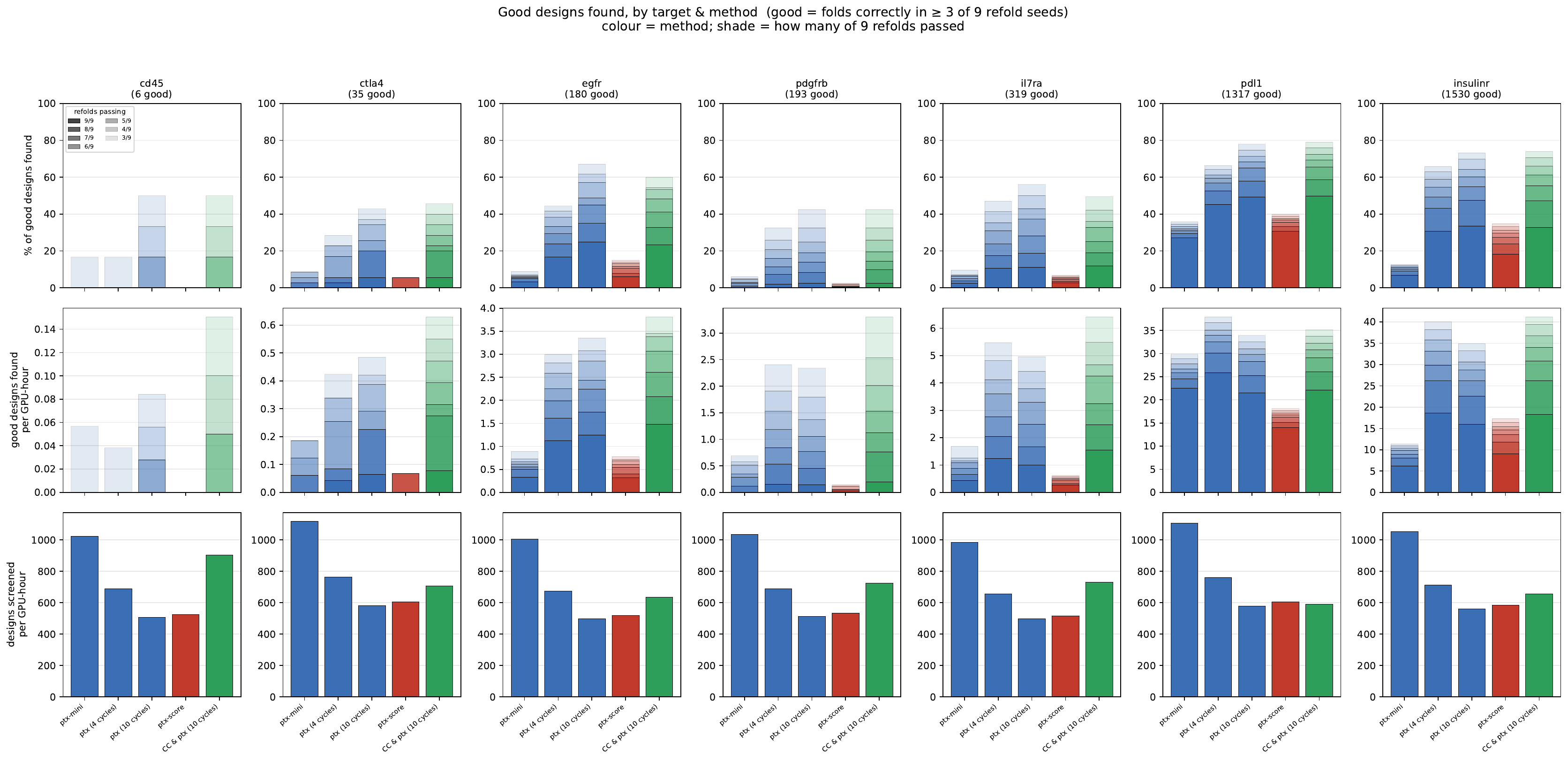}
  \caption{Delivery of reproducible designs by target. A design is
  \emph{good} if it passes at least three of nine held-out refolds;
  seed 101 is reserved for each method's declaration. Targets are
  ordered by their number of good designs. Colour identifies the
  method and shade the number of passing jury seeds.
  \textbf{Top:} recall among good designs.
  \textbf{Middle:} good designs delivered per GPU-hour.
  \textbf{Bottom:} designs screened per GPU-hour.
  The gate delivers more good designs per GPU-hour than ptx@10
  on all seven targets, with gains from $3.5\%$ (PD-L1) to
  $78.5\%$ (CD45), despite lower recall on EGFR and IL7RA.}
  \label{fig:per-target}
\end{figure}

\begin{figure}[ht]
  \centering
  \includegraphics[width=0.9\textwidth]{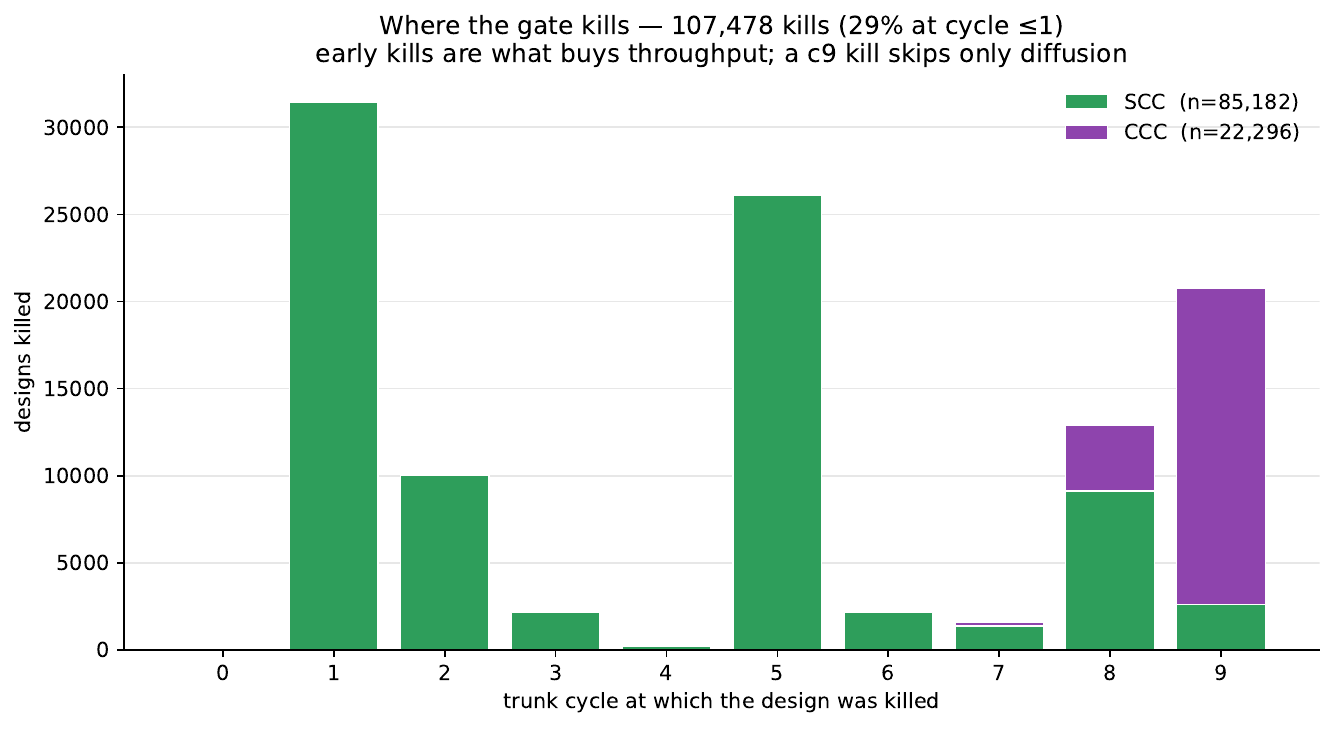}
  \caption{Exit decisions of the union gate. Bars show terminations
  by zero-based trunk-cycle index and the gate that fired. 
  Of approximately $126{,}000$ designs, $107{,}478$
  ($85.6\%$) are terminated: $85{,}182$ by sCC and $22{,}296$
  by cCC. Their median exit indices are $3$ and $9$,
  corresponding to four and ten completed trunk passes.}
  \label{fig:kills-by-cycle}
\end{figure}

\subsection{Simulating gating performance on a smaller set of generated designs}

\begin{figure}[ht]
  \centering
  \includegraphics[width=\linewidth]{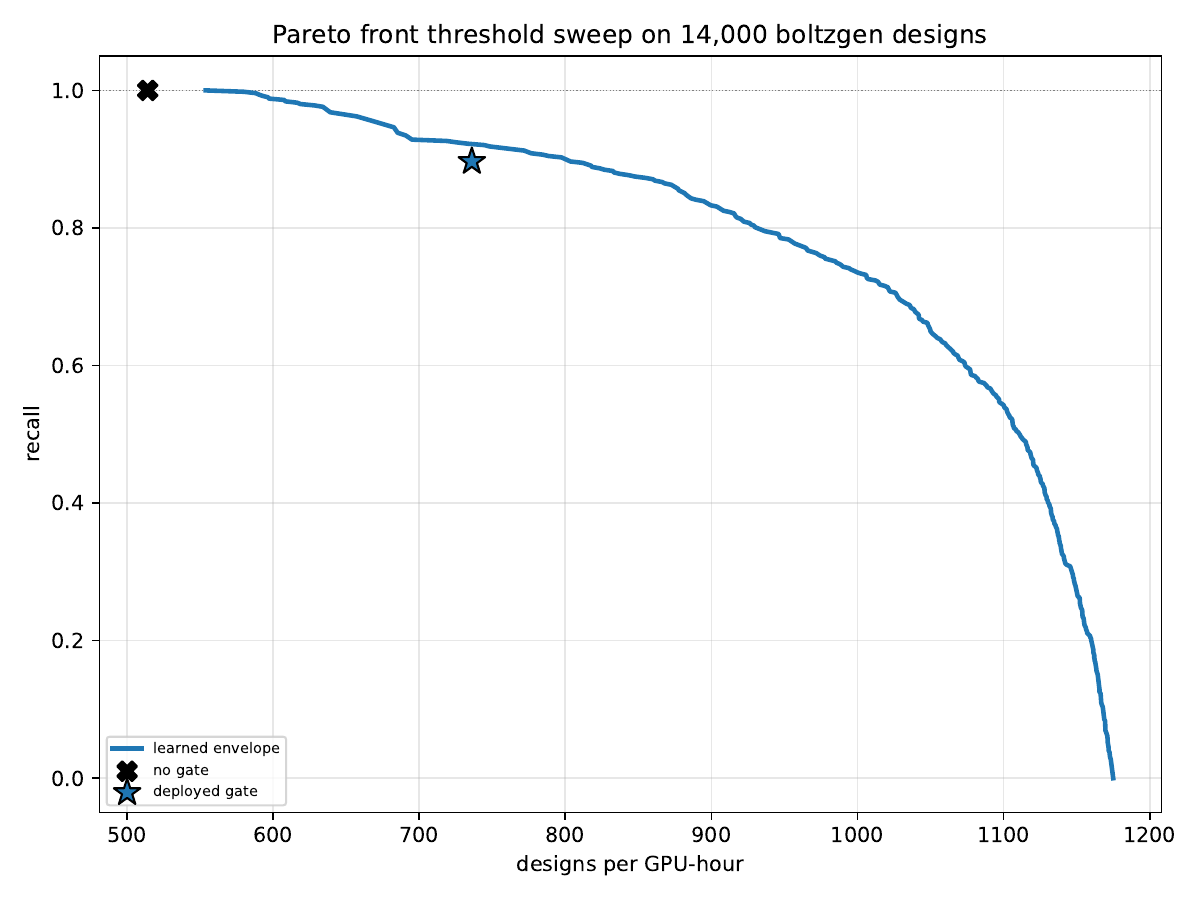}
  \caption{Simulated trade-off between pre-diffusion recall and
  screening throughput on $14{,}000$ generated designs.
  Recall uses the $\geq3/9$ held-out-refold label. The star marks
  the deployed thresholds, which were selected independently of
  this evaluation set.}
  \label{fig:v3-pareto}
\end{figure}

We evaluated the critics on a separate set of $14{,}000$ generated
designs, with $2{,}000$ per target. Sweeping their per-cycle thresholds
gives the trade-off between pre-diffusion retention of designs passing
at least three of nine held-out refolds and screening throughput.
Note that the deployed thresholds in the full experiment were selected independently of this evaluation
set. \Cref{fig:v3-pareto} shows the resulting frontier and the deployed
operating point.

\paragraph{Cost model.}
For a design, modeled runtime in seconds is
\[
  \mathrm{cost}=F_0+cC+sD,
\]
where $F_0$ is a fixed per-design cost ($2.76$\,s gated,
$2.23$\,s ungated), $c$ is the number of completed trunk passes,
$C=0.30$\,s is the cost per pass, $s$ indicates survival to diffusion,
and $D=1.74$\,s is the diffusion cost. A design terminated at
zero-based cycle index $k$ has $c=k+1$; a survivor has $c=10$.
Throughput is $3600/\overline{\mathrm{cost}}$ designs per GPU-hour.
The printed constants give an ungated cost of approximately $7$\,s
per design, or approximately $514$ designs per GPU-hour using the
unrounded fitted values.

\begin{table}[t]
  \centering
  \begin{tabular}{llccc}
    \toprule
    Gate & Critics & Designs/GPU-hour & Recall ($\geq3/9$) & Kill rate \\
    \midrule
    No gate & --- & 514.2 & 1.0000 & 0.000 \\
    Learned (deployed) & cCC $\cup$ sCC & \textbf{736.0}
      & \textbf{0.8966} & 0.848 \\
    Analytic (deploy-like) & ipTM proxy $\cup$ dCC
      & 721.6 & 0.8549 & 0.946 \\
    \bottomrule
  \end{tabular}
  \caption{Simulated prescreening on $14{,}000$ VHH designs.
  Recall is the fraction of $\geq3/9$ designs retained before
  diffusion; it is not final declaration recall. Critics use
  their deployed per-cycle thresholds and minimum-cycle guards.
  Seed 101 is excluded from the nine-seed reference label.}
  \label{tab:prescreen-gates}
\end{table}

\section{Design with ESMFold2}
\label{app:ESM-design}

\subsection{Base ESMFold2 design objective}

We build on the gradient-guided binder design procedure of the ESMFold2 paper,
using their two folding trunks, \texttt{ESMFold2-Experimental-Fast} (cutoff 2021)
and \texttt{ESMFold2-Experimental-Fast-Cutoff2025}, with one trunk randomly sampled each optimization step.
The base structure loss $\mathcal{L}_{\mathrm{struct}}$ is computed entirely from the predicted distogram,
\begin{equation}
  \mathcal{L}_{\mathrm{struct}}
    = 0.5\,\mathcal{L}_{\mathrm{intra}}
    + 0.5\,\mathcal{L}_{\mathrm{inter}}
    + 0.2\,\mathcal{L}_{\mathrm{glob}},
\end{equation}
where $\mathcal{L}_{\mathrm{intra}}$ and $\mathcal{L}_{\mathrm{inter}}$ are
distogram-based intra-binder and binder--target contact terms, and
$\mathcal{L}_{\mathrm{glob}}$ is a radius-of-gyration globularity term. 
A sequence-plausibility regularizer (ESM-C pseudo-perplexity) is applied identically in the base and CycleCritic runs and combined at the gradient level, normalized per design, with weight $0.05$.
Notably, the base objective contains no differentiable structure of confidence-head based term: for design they also skip diffusion and confidence head.

\subsection{CycleCritic-augmented objective}

Since the original ESM-design randomly picks one of the two experimental models per step for generation, we trained different CCs to predict each of their pair representations.

Further we compute two CycleCritic-based losses 
\begin{align}
  \mathcal{L}^{(t)}_{\mathrm{CC\text{-}ipTM}}  &= 1 - \widehat{\mathrm{ipTM}}_{\mathrm{CC}^{(t)}}, \\
  \mathcal{L}^{(t)}_{\mathrm{CC\text{-}ipSAE}} &= 1 - \widehat{\mathrm{ipSAE}}_{\mathrm{CC}^{(t)}}
\end{align}
for $ipTM$ and $\mathrm{ipSAE}_{\mathrm{AG}\rightarrow\mathrm{AB}}$.
Here, each head $CC^{(t)}$ only ever sees $z$ from the trunk $t$ it was trained on. The
full objective is
\begin{equation}
  \mathcal{L}_{\mathrm{struct,CC}}
    = \mathcal{L}_{\mathrm{struct}}
    + 0.5 \left(
        \mathcal{L}^{(t)}_{\mathrm{CC\text{-}ipTM}}
      + \mathcal{L}^{(t)}_{\mathrm{CC\text{-}ipSAE}}
      \right)
\end{equation}

\subsection{Evaluation/Selection metric.}
Like in \citet{biohub2026languagemodeling}, each design is refolded once with each of four checkpoints:
\texttt{ESMFold2-Experimental-Fast},
\texttt{ESMFold2-Experimental-Fast-Cutoff2025},
\texttt{ESMFold2-Experimental}, and
\texttt{ESMFold2-Experimental-Cutoff2025}.
Refolds use three recycling loops and 200 sampling steps,
with the same folding seed shared across checkpoints for
each design. We define the selection score as
\begin{equation}
S_i = \frac{1}{8}\sum_{m=1}^{4}
\left(
\mathrm{ipTM}_{im}
+ \mathrm{ipTM}^{\mathrm{proxy}}_{im}
\right),
\end{equation}

where $\mathrm{ipTM}^{\mathrm{proxy}}$ is the
distogram-derived ipTM surrogate. All designs have complete
scores from all four checkpoints. 

The two Fast checkpoints also serve as generation trunks;
the two non-Fast checkpoints are held out from generation.

\subsection{Guidance parameter study for ESMFold2 binder generation}
\label{app:esm-param-study}

We investigate design-time guidance on four targets
(CD45, EGFR, PDGFR$\beta$, and PD-L1) and three antibody
frameworks (trastuzumab, atezolizumab, and ocankitug),
giving 12 target--framework cells. All nine arms use the same
generator and differ only in their guidance settings.
Each arm contains 255 designs per cell, or 3,060 designs.
The eight guided arms are seed-paired.

Reported errors are within-cell SEs; comparisons use nominal paired $t(11)$ statistics across the 12 cells
\subsubsection{The parameters}
\label{app:param-study:params}

We vary four parameters of the guidance objective and select, axis by axis, the setting that
maximises selection:

\begin{center}
\begin{tabular}{@{}ll@{}}
\toprule
parameter & candidate settings \\
\midrule
guidance source    & none (base), training-free ipTM proxy, CycleCritic \\
critic heads       & ipTM, ipTM $+$ $\mathrm{ipSAE}_{\mathrm{AG}\rightarrow\mathrm{AB}}$ (dual) \\
recycling schedule & all cycles, last cycle only \\
guidance weight    & 0.50, 0.85, 1.00, 1.50 \\
\bottomrule
\end{tabular}
\end{center}

The Cycle-Critic carries per-trunk-matched heads (a critic head trained for the folding trunk it
scores). Unless a comparison holds a parameter fixed to isolate another, arms are compared at the
eventual recommended setting (dual head, all cycles, weight as noted).

\subsubsection{Guidance source --- the learned critic beats the training-free proxy}
\label{app:param-study:source}

Both guidance sources improve substantially over unguided design, and the learned critic beats
the training-free surrogate:

\begin{center}
\begin{tabular}{@{}lrl@{}}
\toprule
guidance source (all cycles, $w$ 0.85) & selection & $\Delta$ vs base \\
\midrule
none (base)                       & 0.7832 & --- \\
ipTM proxy              & 0.8061 & $+0.0229$ ($t(11)=3.4$, 12/12) \\
Cycle-Critic (ipTM)  & 0.8175 & $+0.0343$ ($t(11)=3.3$, 12/12) \\
\bottomrule
\end{tabular}
\end{center}

The critic beats the proxy by a paired $\mathbf{+0.0114}$ (SE 0.0043, $t(11)=2.6$; 10/12 cells),
with the tiers cleanly separated (the weakest critic arm sits above the strongest proxy arm).
Because generator and launcher are held fixed, this is attributable to the guidance objective
alone.

\subsubsection{Critic heads --- the dual head beats ipTM alone}
\label{app:param-study:heads}

Adding a second CC head on top of the ipTM head,
holding routing and weight fixed:

\begin{center}
\begin{tabular}{@{}lrl@{}}
\toprule
critic heads (all cycles, $w$ 0.85) & selection & $\Delta$ \\
\midrule
ipTM            & 0.8175 & --- \\
ipTM $+$ ipSAE  & 0.8267 & $+0.0092$ ($t(11)=1.8$, 8/12) \\
\bottomrule
\end{tabular}
\end{center}

\subsubsection{Recycling schedule --- apply the critic on all cycles}
\label{app:param-study:cycles}

Averaging the guidance over all recycling carry cycles vs.\ applying it only on the final cycle
has a much larger effect for the learned critic than for the proxy:

\begin{center}
\begin{tabular}{@{}lrrcc@{}}
\toprule
guidance source & all $-$ end ($\Delta$ selection) & SE & $t(11)$ & cells won \\
\midrule
learned critic (dual, $w$ 0.85) & $\mathbf{+0.0157}$ & 0.0048 & $\mathbf{3.3}$ & \textbf{12/12} \\
distogram proxy ($w$ 0.85)      & $+0.0021$          & 0.0020 & $1.0$          & 9/12 \\
\bottomrule
\end{tabular}
\end{center}

All-cycle averaging is worth ${\sim}7\times$ more to the learned critic, a clean 12/12 sweep; the
end-cycle dual critic (0.8111) even falls \emph{below} the all-cycle ipTM-only critic (0.8175).
The learned critic evidently extracts usable interface signal from the intermediate recycling
trajectory that the training-free surrogate cannot. 

\subsubsection{Guidance weight --- saturates by about 1.0}
\label{app:param-study:weight}

Sweeping the weight for the dual critic on all cycles (all four points seed-paired):

\begin{center}
\begin{tabular}{@{}crrr@{}}
\toprule
weight & selection & $\Delta$ vs previous & $t(11)$ \\
\midrule
0.50          & 0.8202          & ---       & --- \\
0.85          & 0.8267          & $+0.0065$ & 1.7 (9/12) \\
\textbf{1.00} & \textbf{0.8292} & $+0.0025$ & 1.6 (7/12) \\
1.50          & 0.8308          & $+0.0015$ & 1.2 (8/12) \\
\bottomrule
\end{tabular}
\end{center}

The curve rises monotonically but saturates: the step to 1.00 still adds a measurable increment,
whereas the further step to 1.50 is negligible. This can be explained by
the fact that the gradients of the structure loss get normalized to unit length. Since the gains
after 1.0 are marginal and we want to avoid breaking optimization we choose weight 1.0.

\subsubsection{Recommended configuration}
\label{app:param-study:recommended}

The parameter study converges on \\
\textbf{Cycle-Critic, dual head (ipTM $+$ ipSAE), all recycling
cycles, guidance weight 1.0},\\
 which reaches {0.8292 selection ($+5.88\%$ over unguided
base)}. Full macro-averaged leaderboard over the 12 shared cells ($\pm 1$ SE $=$ within-cell
sampling):

\begin{center}
\begin{tabular}{@{}lrrr@{}}
\toprule
arm & selection & $\pm 1$ SE & $\Delta$ vs base \\
\midrule
Cycle-Critic, dual, all cycles, weight 1.50 & 0.8308 & $\pm0.0012$ & $+6.08\%$ \\
\textbf{Cycle-Critic, dual, all cycles, weight 1.00 --- recommended} & \textbf{0.8292} & $\mathbf{\pm0.0013}$ & $\mathbf{+5.88\%}$ \\
Cycle-Critic, dual, all cycles, weight 0.85 & 0.8267 & $\pm0.0013$ & $+5.56\%$ \\
Cycle-Critic, dual, all cycles, weight 0.50 & 0.8202 & $\pm0.0014$ & $+4.73\%$ \\
Cycle-Critic, ipTM, all cycles, weight 0.85 & 0.8175 & $\pm0.0014$ & $+4.39\%$ \\
Cycle-Critic, dual, end cycle, weight 0.85  & 0.8111 & $\pm0.0014$ & $+3.56\%$ \\
iptm-proxy, all cycles, weight 0.85         & 0.8061 & $\pm0.0016$ & $+2.93\%$ \\
iptm-proxy, end cycle, weight 0.85          & 0.8040 & $\pm0.0016$ & $+2.66\%$ \\
base (no guidance)                          & 0.7832 & $\pm0.0018$ & --- \\
\bottomrule
\end{tabular}
\end{center}

Summarizing, while the iptm-proxy already improves hallucination-based design ($+2.93\%$ over
base), the Cycle-Critic wins out for two reasons: it is flexible enough to model other confidence
metrics (the added ipSAE head), and its early-cycle predictions are far more informative than the
proxy's --- guiding on all recycling cycles rather than only the last helps the critic roughly
seven times as much as it helps the proxy. The recommended configuration reaches $+5.88\%$ over
base, roughly double the proxy's gain.

\subsection{Results}

While the additional compute of the two loss terms across the two CycleCritic executions
leads to a modest increase of $0.4\%$ in time per design (\Cref{tab:cc-overhead}), the
optimization metric used by ESMFold2-design favors the CC-based optimization in 6 out
of 7 targets for both VHHs and Fvs (\Cref{tab:esm_6h-pooled-topk}). The optimization
metric is the mean of the ipTM and distogram proxy predicted by four different critics
refolding the designed sequence.

\begin{table}[t]
\centering
\setlength{\tabcolsep}{3.5pt}
\caption{Design-suite ESM results pooled by framework class and target, at equal 6 GPU-h per cell. Metric is mean4(ipTM, ipTM distogram proxy). Top-$K$ columns report the mean of a cell's $K$ best designs. The budget is drawn against the critic: the measured $+0.4\%$ per-design overhead is applied, so CycleCritic is allotted slightly fewer designs than the baseline. Larger Base/CC entries are bold; the summary row gives the format totals and the mean across targets.}
\label{tab:esm_6h-pooled-topk}
\resizebox{\linewidth}{!}{%
\begin{tabular}{llc|cc|cc|cc|cc|cc|cc}
\toprule
\multirow{2}{*}{Format} & \multirow{2}{*}{Target} & \multirow{2}{*}{$n$ (Base/CC)}
& \multicolumn{2}{c}{Mean}
& \multicolumn{2}{c}{Median}
& \multicolumn{2}{c}{Top-5}
& \multicolumn{2}{c}{Top-10}
& \multicolumn{2}{c}{Top-25}
& \multicolumn{2}{c}{Top-50} \\
\cmidrule(lr){4-5}\cmidrule(lr){6-7}\cmidrule(lr){8-9}\cmidrule(lr){10-11}\cmidrule(lr){12-13}\cmidrule(lr){14-15}
& & & Base & CC & Base & CC & Base & CC & Base & CC & Base & CC & Base & CC \\
\midrule
\multirow{8}{*}{VHH}
& CD45     & $403/401$ & $.522$ & $\mathbf{.739}$ & $.477$ & $\mathbf{.790}$ & $.914$ & $\mathbf{.929}$ & $.908$ & $\mathbf{.925}$ & $.895$ & $\mathbf{.919}$ & $.855$ & $\mathbf{.913}$ \\
& CTLA-4   & $600/596$ & $.846$ & $\mathbf{.860}$ & $.876$ & $\mathbf{.883}$ & $\mathbf{.919}$ & $.917$ & $\mathbf{.917}$ & $.916$ & $.912$ & $\mathbf{.913}$ & $.908$ & $\mathbf{.911}$ \\
& EGFR     & $392/392$ & $.762$ & $\mathbf{.817}$ & $.798$ & $\mathbf{.845}$ & $.913$ & $\mathbf{.918}$ & $.905$ & $\mathbf{.914}$ & $.894$ & $\mathbf{.908}$ & $.883$ & $\mathbf{.900}$ \\
& IL7RA    & $388/388$ & $\mathbf{.748}$ & $.669$ & $\mathbf{.783}$ & $.678$ & $\mathbf{.920}$ & $.905$ & $\mathbf{.915}$ & $.899$ & $\mathbf{.909}$ & $.888$ & $\mathbf{.900}$ & $.873$ \\
& INSR     & $477/476$ & $.812$ & $\mathbf{.856}$ & $.841$ & $\mathbf{.878}$ & $.928$ & $\mathbf{.932}$ & $.924$ & $\mathbf{.929}$ & $.916$ & $\mathbf{.925}$ & $.909$ & $\mathbf{.920}$ \\
& PDGFR$\beta$ & $400/400$ & $.762$ & $\mathbf{.792}$ & $.811$ & $\mathbf{.818}$ & $.896$ & $\mathbf{.900}$ & $.893$ & $\mathbf{.897}$ & $.887$ & $\mathbf{.890}$ & $.878$ & $\mathbf{.883}$ \\
& PD-L1    & $604/600$ & $.855$ & $\mathbf{.875}$ & $.883$ & $\mathbf{.895}$ & $\mathbf{.947}$ & $.945$ & $\mathbf{.944}$ & $.940$ & $\mathbf{.939}$ & $.936$ & $\mathbf{.935}$ & $.933$ \\
\cmidrule(lr){2-15}
& \textit{mean} & $3264/3253$ & $.758$ & $\mathbf{.801}$ & $.781$ & $\mathbf{.827}$ & $.919$ & $\mathbf{.921}$ & $.915$ & $\mathbf{.917}$ & $.907$ & $\mathbf{.911}$ & $.896$ & $\mathbf{.905}$ \\
\midrule
\multirow{8}{*}{Fv}
& CD45     & $159/158$ & $.631$ & $\mathbf{.793}$ & $.616$ & $\mathbf{.831}$ & $.875$ & $\mathbf{.890}$ & $.869$ & $\mathbf{.884}$ & $.851$ & $\mathbf{.876}$ & $.821$ & $\mathbf{.867}$ \\
& CTLA-4   & $222/221$ & $.860$ & $\mathbf{.864}$ & $.870$ & $\mathbf{.870}$ & $\mathbf{.910}$ & $.907$ & $\mathbf{.906}$ & $.904$ & $\mathbf{.899}$ & $.899$ & $.894$ & $\mathbf{.894}$ \\
& EGFR     & $156/156$ & $.829$ & $\mathbf{.852}$ & $.859$ & $\mathbf{.874}$ & $\mathbf{.921}$ & $.921$ & $.916$ & $\mathbf{.918}$ & $.906$ & $\mathbf{.910}$ & $.897$ & $\mathbf{.902}$ \\
& IL7RA    & $156/155$ & $.775$ & $\mathbf{.778}$ & $\mathbf{.812}$ & $.805$ & $\mathbf{.893}$ & $.892$ & $\mathbf{.886}$ & $.883$ & $\mathbf{.874}$ & $.869$ & $\mathbf{.862}$ & $.854$ \\
& INSR     & $183/181$ & $.806$ & $\mathbf{.832}$ & $.828$ & $\mathbf{.849}$ & $.907$ & $\mathbf{.910}$ & $.899$ & $\mathbf{.904}$ & $.888$ & $\mathbf{.896}$ & $.879$ & $\mathbf{.889}$ \\
& PDGFR$\beta$ & $159/159$ & $.780$ & $\mathbf{.791}$ & $.789$ & $\mathbf{.805}$ & $\mathbf{.869}$ & $.862$ & $\mathbf{.862}$ & $.859$ & $.851$ & $\mathbf{.852}$ & $.838$ & $\mathbf{.841}$ \\
& PD-L1    & $225/225$ & $.879$ & $\mathbf{.882}$ & $.891$ & $\mathbf{.892}$ & $\mathbf{.934}$ & $.929$ & $\mathbf{.930}$ & $.926$ & $\mathbf{.923}$ & $.921$ & $\mathbf{.917}$ & $.917$ \\
\cmidrule(lr){2-15}
& \textit{mean} & $1260/1255$ & $.794$ & $\mathbf{.828}$ & $.809$ & $\mathbf{.847}$ & $.901$ & $\mathbf{.902}$ & $.895$ & $\mathbf{.897}$ & $.885$ & $\mathbf{.889}$ & $.873$ & $\mathbf{.881}$ \\
\bottomrule
\end{tabular}%
}
\end{table}

\begin{table}[t]
  \centering
  \caption{Cycle-Critic (dual ipTM+ipSAE) generation-time for ESM hallucination.
    The per-design row is the 149-step optimization loop}
  \label{tab:cc-overhead}
  \begin{tabular}{llrrr}
    \toprule
    Quantity & Format & Base & CC & $\Delta$ \\
    \midrule
    \multirow{2}{*}{Per design}
      & VHH & $130.8$\,s & $131.3$\,s & $+0.4\%$ \\
      & Fv  & $265.9$\,s & $267.0$\,s & $+0.4\%$ \\
    \bottomrule
  \end{tabular}
\end{table}

\section{Hallucination with Mosaic}
\label{App:Mosaic}

We compare Mosaic's \citep{boyd2025mosaic} standard differentiable sequence design with full Protenix against a
trunk-only CycleCritic objective. 
The purpose of the CycleCritic objective is to replace the expensive diffusion and confidence-head losses during optimization.
Final selected designs were still evaluated by full Protenix refolding model.

\subsection{Original mosaic objective}

The standard Mosaic objective we used in the baseline design runs is
\begin{align}
\mathcal{L}_{\mathrm{full}}
={}&
\mathcal{L}_{\mathrm{contact}}
+0.05\,\mathcal{L}_{\mathrm{target\to binder\ PAE}}
+0.05\,\mathcal{L}_{\mathrm{binder\to target\ PAE}} \notag \\
&+0.025\,\mathcal{L}_{\mathrm{iPTM}}
+0.4\,\mathcal{L}_{\mathrm{binder\ PAE}}
+0.025\,\mathcal{L}_{\mathrm{pTM\text{-}energy}}
+0.1\,\mathcal{L}_{\mathrm{pLDDT}} .
\end{align}

Here, $\mathcal{L}_{\mathrm{contact}}$ is a distogram-based binder-target
contact term. The two interface PAE terms penalize predicted aligned error from
target to binder and binder to target. The binder PAE term penalizes predicted
aligned error within the designed binder chain. 
The ipTM and pLDDT terms come from the full Protenix confidence machinery, while
$\mathcal{L}_{\mathrm{pTM\text{-}energy}}$ is a pTM-like energy computed from the
PAE logits.

\subsection{CycleCritic replacement objective}

In the CycleCritic runs, we replaced the confidence-dependent terms by lightweight CycleCritic predictions computed from the pair representations.
Our objective then was
\begin{align}
\mathcal{L}_{\mathrm{CC}}
={}&
\mathcal{L}_{\mathrm{contact}}
+0.05\,\mathcal{L}_{\mathrm{target\to binder\ dist}}
+0.05\,\mathcal{L}_{\mathrm{binder\to target\ dist}} \notag \\
&+0.025\,\mathcal{L}_{\mathrm{CC\text{-}iPTM}}
+0.4\,\mathcal{L}_{\mathrm{binder\ dist}}
+0.025\,\mathcal{L}_{\mathrm{CC\text{-}pTM}}
+0.1\,\mathcal{L}_{\mathrm{CC\text{-}pLDDT}} .
\end{align}

The pTM-energy term was replaced by the pTM CycleCritic prediction in the same
weighted slot. 

The CycleCritic metric losses are defined as
\begin{align}
\mathcal{L}_{\mathrm{CC\text{-}iPTM}} &= -\widehat{\mathrm{iPTM}}_{\mathrm{CC}},\\
\mathcal{L}_{\mathrm{CC\text{-}pTM}} &= -\widehat{\mathrm{pTM}}_{\mathrm{CC}},\\
\mathcal{L}_{\mathrm{CC\text{-}pLDDT}} &= -\widehat{\mathrm{pLDDT}}_{\mathrm{CC}}.
\end{align}

Each CycleCritic head predicts a distribution over bins for the corresponding
confidence metric, and the scalar prediction is the expected value of that
distribution:
\begin{equation}
\widehat{m}_{\mathrm{CC}}
=
\sum_b p_{\mathrm{CC}}(b)\,c_b,
\qquad
m\in\{\mathrm{iPTM},\mathrm{pTM},\mathrm{pLDDT}\}.
\end{equation}

For runs using all trunk cycles, the CycleCritic predictions were averaged over
cycles:
\begin{equation}
\widehat{m}_{\mathrm{CC}}
=
\frac{1}{C}\sum_{c=1}^{C}\widehat{m}_{\mathrm{CC}}^{(c)} .
\end{equation}

The PAE terms in the original full-Protenix objective were replaced by
distogram expected-distance terms. For each residue pair, the expected distance
was computed from the distogram distribution:
\begin{equation}
\mathbb{E}[d_{ij}]
=
\sum_b p_{ij}(b)\,d_b .
\end{equation}

The binder-target distance term averages the nearest expected distance to the target over binder residues:
\begin{equation}
\mathcal{L}_{\mathrm{binder\to target\ dist}}
=
\frac{1}{N_{\mathrm{binder}}}
\sum_{i\in\mathrm{binder}}
\min_{j\in\mathrm{target}}\mathbb{E}[d_{ij}] .
\end{equation}

The binder-internal distance term averages expected distances within the binder:
\begin{equation}
\mathcal{L}_{\mathrm{binder\ dist}}
=
\frac{1}{N_{\mathrm{binder}}^2}
\sum_{i,j\in\mathrm{binder}}\mathbb{E}[d_{ij}] .
\end{equation}

We did not apply hyperparameter optimization to fine-tune the loss weights.

\subsection{Design objectives and optimization settings}

Both objectives used identical optimization settings. We used APGM with a
three-stage schedule totaling $50+30+30=110$ optimization steps:
\begin{center}
\begin{tabular}{lccccc}
\toprule
Stage & Steps & Step size & Momentum & Scale & Log-space \\
\midrule
1 & 50 & $1.5\sqrt{N_{\mathrm{design}}}$ & 0.2 & 1.0 & yes \\
2 & 30 & $0.5\sqrt{N_{\mathrm{design}}}$ & 0.0 & 1.1 & no \\
3 & 30 & $0.25\sqrt{N_{\mathrm{design}}}$ & 0.0 & 1.1 & yes \\
\bottomrule
\end{tabular}
\end{center}

The production runs used batch size 1, two Protenix recycling steps during
optimization, one final sample per seed, and full Protenix-v2 refolding with 10 cycles and 200 diffusion steps for endpoint evaluation.

We ran the optimization for 6h per target $\times$ framework and loss, so 49 combinations to compare on.
\subsection{Results}

The results can be seen in \Cref{tab:mosaic_hq-pooled-topk}.
The CC speeds up sampling substantially, while yielding a similar metric distribution.

\begin{table}[t]
\centering
\setlength{\tabcolsep}{3.5pt}
\caption{Design-suite Mosaic results pooled by framework class and target, at equal GPU-time. Metric is held-out full-Protenix refold ipTM. Top-$K$ columns report the mean of a cell's $K$ best designs, which is rank-comparable across arms holding unequal numbers of designs. Larger Base/CC entries are bold. $n$ is the number of designs each arm produced for that target within the shared budget; the summary row gives the format totals and the mean across targets.}
\label{tab:mosaic_hq-pooled-topk}
\resizebox{\linewidth}{!}{%
\begin{tabular}{llc|cc|cc|cc|cc|cc|cc}
\toprule
\multirow{2}{*}{Format} & \multirow{2}{*}{Target} & \multirow{2}{*}{$n$ (Base/CC)}
& \multicolumn{2}{c}{Mean}
& \multicolumn{2}{c}{Median}
& \multicolumn{2}{c}{Top-5}
& \multicolumn{2}{c}{Top-10}
& \multicolumn{2}{c}{Top-25}
& \multicolumn{2}{c}{Top-50} \\
\cmidrule(lr){4-5}\cmidrule(lr){6-7}\cmidrule(lr){8-9}\cmidrule(lr){10-11}\cmidrule(lr){12-13}\cmidrule(lr){14-15}
& & & Base & CC & Base & CC & Base & CC & Base & CC & Base & CC & Base & CC \\
\midrule
\multirow{8}{*}{VHH}
& CD45     & $175/282$ & $.363$ & $\mathbf{.372}$ & $\mathbf{.291}$ & $.281$ & $.873$ & $\mathbf{.905}$ & $.819$ & $\mathbf{.888}$ & $.738$ & $\mathbf{.821}$ & $.624$ & $\mathbf{.744}$ \\
& CTLA-4   & $253/497$ & $\mathbf{.431}$ & $.427$ & $\mathbf{.353}$ & $.337$ & $.899$ & $\mathbf{.917}$ & $.884$ & $\mathbf{.911}$ & $.843$ & $\mathbf{.891}$ & $.784$ & $\mathbf{.860}$ \\
& EGFR     & $167/253$ & $.339$ & $\mathbf{.353}$ & $.294$ & $\mathbf{.313}$ & $.814$ & $\mathbf{.837}$ & $.743$ & $\mathbf{.783}$ & $.620$ & $\mathbf{.685}$ & $.518$ & $\mathbf{.589}$ \\
& IL7RA    & $165/263$ & $.428$ & $\mathbf{.446}$ & $.383$ & $\mathbf{.395}$ & $.887$ & $\mathbf{.891}$ & $.841$ & $\mathbf{.880}$ & $.759$ & $\mathbf{.841}$ & $.678$ & $\mathbf{.780}$ \\
& INSR & $191/313$ & $.521$ & $\mathbf{.532}$ & $.477$ & $\mathbf{.497}$ & $\mathbf{.911}$ & $.904$ & $.883$ & $\mathbf{.890}$ & $.822$ & $\mathbf{.863}$ & $.767$ & $\mathbf{.826}$ \\
& PDGFR$\beta$   & $168/269$ & $.432$ & $\mathbf{.448}$ & $.397$ & $\mathbf{.421}$ & $.882$ & $\mathbf{.898}$ & $.843$ & $\mathbf{.866}$ & $.742$ & $\mathbf{.805}$ & $.647$ & $\mathbf{.739}$ \\
& PD-L1    & $241/440$ & $.526$ & $\mathbf{.557}$ & $.491$ & $\mathbf{.531}$ & $.930$ & $\mathbf{.934}$ & $.920$ & $\mathbf{.930}$ & $.891$ & $\mathbf{.918}$ & $.827$ & $\mathbf{.893}$ \\
\cmidrule(lr){2-15}
& \textit{mean} & $1360/2317$ & $.434$ & $\mathbf{.448}$ & $.384$ & $\mathbf{.397}$ & $.885$ & $\mathbf{.898}$ & $.848$ & $\mathbf{.878}$ & $.773$ & $\mathbf{.832}$ & $.692$ & $\mathbf{.776}$ \\
\midrule
\multirow{8}{*}{FV}
& CD45     & $82/106$ & $\mathbf{.295}$ & $.294$ & $\mathbf{.231}$ & $.231$ & $.667$ & $\mathbf{.725}$ & $.617$ & $\mathbf{.678}$ & $.506$ & $\mathbf{.561}$ & $.376$ & $\mathbf{.426}$ \\
& CTLA-4   & $107/154$ & $\mathbf{.335}$ & $.329$ & $\mathbf{.297}$ & $.279$ & $.630$ & $\mathbf{.787}$ & $.578$ & $\mathbf{.713}$ & $.501$ & $\mathbf{.588}$ & $.420$ & $\mathbf{.478}$ \\
& EGFR     & $81/98$ & $.353$ & $\mathbf{.366}$ & $\mathbf{.305}$ & $.298$ & $.746$ & $\mathbf{.758}$ & $.659$ & $\mathbf{.716}$ & $.539$ & $\mathbf{.622}$ & $.431$ & $\mathbf{.489}$ \\
& IL7RA    & $78/100$ & $\mathbf{.337}$ & $.336$ & $\mathbf{.311}$ & $.293$ & $.675$ & $\mathbf{.713}$ & $.625$ & $\mathbf{.674}$ & $.523$ & $\mathbf{.581}$ & $.423$ & $\mathbf{.459}$ \\
& INSR & $85/115$ & $\mathbf{.526}$ & $.500$ & $\mathbf{.521}$ & $.461$ & $.839$ & $\mathbf{.854}$ & $.803$ & $\mathbf{.837}$ & $.736$ & $\mathbf{.773}$ & $.645$ & $\mathbf{.663}$ \\
& PDGFR$\beta$   & $78/100$ & $\mathbf{.342}$ & $.336$ & $\mathbf{.302}$ & $.292$ & $.684$ & $\mathbf{.708}$ & $.619$ & $\mathbf{.647}$ & $.501$ & $\mathbf{.558}$ & $.407$ & $\mathbf{.452}$ \\
& PD-L1    & $101/142$ & $\mathbf{.458}$ & $.392$ & $\mathbf{.402}$ & $.325$ & $.837$ & $\mathbf{.861}$ & $\mathbf{.814}$ & $.814$ & $\mathbf{.743}$ & $.725$ & $\mathbf{.614}$ & $.597$ \\
\cmidrule(lr){2-15}
& \textit{mean} & $612/815$ & $\mathbf{.378}$ & $.365$ & $\mathbf{.339}$ & $.311$ & $.725$ & $\mathbf{.772}$ & $.674$ & $\mathbf{.725}$ & $.578$ & $\mathbf{.630}$ & $.474$ & $\mathbf{.509}$ \\
\bottomrule
\end{tabular}%
}
\end{table}

\begin{table}[t]
\centering
\caption{Measured per-design wall-clock generation time for the Mosaic campaign at equal GPU-time (six hours per cell and arm). Duration is the gap between consecutive design start-times within a worker, so it captures designs per unit GPU-time including per-design overhead. $\Delta$ is the change of CycleCritic relative to the baseline.}
\label{tab:mosaic-runtime}
\begin{tabular}{lccc|ccc}
\toprule
\multirow{2}{*}{Format} & \multicolumn{3}{c}{Mean (s)} & \multicolumn{3}{c}{Median (s)} \\
\cmidrule(lr){2-4}\cmidrule(lr){5-7}
& Base & CC & $\Delta$ & Base & CC & $\Delta$ \\
\midrule
VHH & $453.2$ & $261.4$ & $\mathbf{-42.3\%}$ & $465.0$ & $262.0$ & $\mathbf{-43.7\%}$ \\
FV & $790.7$ & $588.5$ & $\mathbf{-25.6\%}$ & $795.0$ & $582.5$ & $\mathbf{-26.7\%}$ \\
\midrule
All & $552.1$ & $341.6$ & $\mathbf{-38.1\%}$ & $515.5$ & $308.0$ & $\mathbf{-40.3\%}$ \\
\bottomrule
\end{tabular}%
\end{table}

\section{Analysis of binder designs}
\label{app:analysis_campaign}


Analyses were run over the full generated pool of both campaigns, across two formats (VHH, scFv) and seven antigens. Comparisons are within a campaign and format. All quantities are reported as rates or per-design means rather than raw totals, as the equal-time  budget for the different methods leave the baseline and CycleCritic arms with unequal pool sizes.

\subsection{Diversity and novelty}

\paragraph{Within-sample diversity.} Designs are compared on their concatenated designed CDRs. The exact-duplicate rate is zero everywhere, with a unique fraction of $1.000$ in all cells in both baseline and CycleCritic arms. Pairwise CDR-H3 identity is computed with a Levenshtein-indel normalised similarity within each cell and sits at ${\sim}27$--$30\%$ in every condition, with a CycleCritic-versus-baseline difference of at most $0.6$\% (\Cref{fig:CC-design}a). Greedy clustering of the same identity matrix at $70$, $80$ and $90\%$ leaves approximately one design per cluster throughout, so no near-duplicate families form in either arm.

\paragraph{Novelty against the human repertoire.} For every design we locate its closest natural counterpart among $2{,}295{,}779$ unique human heavy CDR-H3 loops drawn from $3.0$M paired antibodies across $19$ OAS studies, matching on length and scoring by Hamming identity. No design reproduces a natural CDR-H3 exactly ($0$ of $14,136$), and the closest human loop is only ${\sim}41$--$48\%$ identical (\Cref{fig:CC-design}b).  CycleCritic is equal or more novel than baseline in every cell. Importantly, since paired OAS contains no camelid sequences, nanobody CDR-H3s are scored against human heavy chains, inflating their apparent novelty by construction; both arms are scored against the same reference, however, so the comparison between them is unaffected.

\paragraph{Antibody language-model likelihood and humanness.} We assess the designs' naturalness and humanness with two models that are not part of the design loop, AbLang2 pseudo-log-likelihood and the OASis humanness score. By equivalence testing, CycleCritic is equal or better in all eight comparisons (\Cref{fig:CC-design}c), with uniformly small effect sizes, suggesting CycleCritic does not degrade designs naturalness.

\paragraph{Interface diversity.} Sequence diversity does not preclude designs converging on a common binding mode, so we compare interfaces directly with iAlign, scoring all pairs within each (framework, target, arm) cell of both campaigns. Mean $1 - \mathrm{IS}$ is $0.753$ for baseline and $0.748$ for CycleCritic in the ESMFold2 campaign, and $0.831$ against $0.821$ in Mosaic, indicating low similarity (\Cref{fig:CC-design}d).

\begin{figure}
    \centering
    \includegraphics[width=\linewidth]{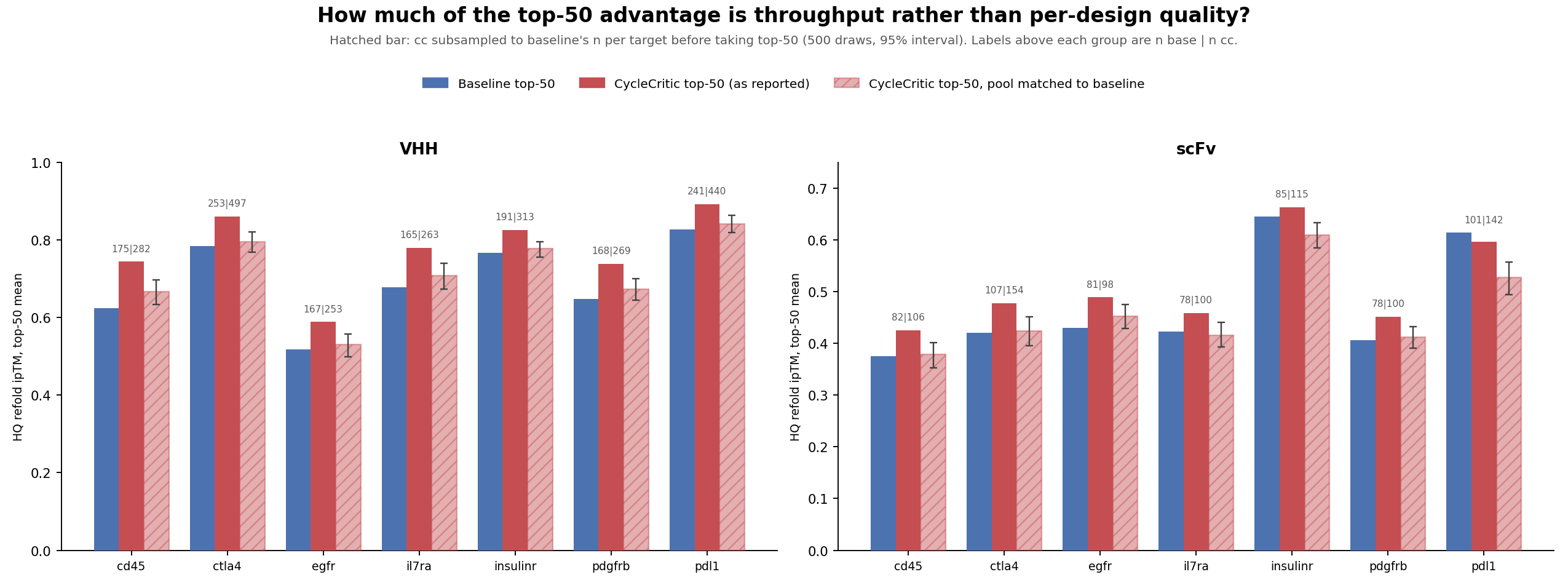}
    \caption{\textbf{How much of the Mosaic top-50 advantage is throughput rather than per-design quality.} Per antigen, baseline top-50 (blue), CycleCritic top-50 as reported (red), and CycleCritic top-50 after subsampling its pool to the baseline's $n$ before taking the top-50 (hatched, 500 draws, 95\% interval). Labels above each group are $n$ baseline $|$ $n$ CycleCritic. Where the hatched bar falls back toward the baseline the top-50 edge is largely a more-shots effect; where it stays near the solid red bar the edge survives pool-matching.}
    \label{fig:CC-throughput}
\end{figure}

\paragraph{Throughput and the top-K advantage.} In the ESMFold2 campaign CycleCritic costs $+0.4\%$ per design and lifts the whole distribution. In Mosaic its per-design optimisation is $\sim\!1.6\times$ faster, so the same $6$ GPU-h/cell budget yields $1{,}972$ baseline against $3{,}132$ CycleCritic designs, and the advantage is concentrated in the selected top-K rather than the full-set mean. Subsampling the CycleCritic pool to the baseline's $n$ before taking each arm's top-50 (\Cref{fig:CC-throughput}) shows how much of that edge is due to throughput. This further supports that the underlying methods are not degraded by CycleCritic; instead, CycleCritic provides a speedup that allows us to produce more designs in less time, improving the best-of-$K$ selection under a fixed time budget.

\end{document}

%% file: math_commands.tex
\usepackage{amsmath,amsfonts,bm}

\def\eqref#1{equation~\ref{#1}}

\def\1{\bm{1}}

\DeclareMathAlphabet{\mathsfit}{\encodingdefault}{\sfdefault}{m}{sl}
\SetMathAlphabet{\mathsfit}{bold}{\encodingdefault}{\sfdefault}{bx}{n}



%% file: poppy-CC-scheme.tex
\definecolor{arrowcolor}{RGB}{0,0,0}
\definecolor{embedfill}{RGB}{238,242,255}
\definecolor{embedstroke}{RGB}{79,70,229}
\definecolor{cardborder}{RGB}{90,105,130}
\definecolor{pairstroke}{RGB}{13,148,136}
\definecolor{singlestroke}{RGB}{234,88,12}

\colorlet{trunkfill}{embedfill}
\colorlet{trunkstroke}{embedstroke}

\definecolor{criticfill}{RGB}{255,241,242}
\definecolor{criticstroke}{RGB}{225,29,72}

\colorlet{structfill}{embedfill}
\colorlet{structstroke}{embedstroke}
\colorlet{conffill}{embedfill}
\colorlet{confstroke}{embedstroke}

\definecolor{containerbg}{RGB}{248,250,252}
\definecolor{containerstroke}{RGB}{148,163,184}
\definecolor{checkgreen}{RGB}{22,163,74}
\definecolor{crossred}{RGB}{220,38,38}

\tikzset{
    softshadow/.style={
        drop shadow={
            shadow xshift=1.5pt,
            shadow yshift=-2pt,
            opacity=0.12
        }
    },
    rblock/.style={
        thick,
        align=center,
        font=\sffamily\bfseries\scriptsize,
        softshadow,
        draw=#1stroke,
        fill=#1fill,
        text=#1stroke!90!black,
        rounded corners=6pt,
        minimum width=1.8cm,
        minimum height=1.1cm
    },
    card/.style={
        very thick,
        draw=cardborder,
        fill=white,
        rounded corners=9pt,
        softshadow
    },
    decision/.style={
        draw=containerstroke,
        fill=white,
        thick,
        rounded corners=8pt,
        font=\sffamily\bfseries\tiny,
        inner xsep=6pt,
        inner ysep=3.5pt
    },
    gate/.style={
        diamond,
        draw=containerstroke,
        fill=white,
        thick,
        aspect=1.5,
        inner sep=2pt,
        font=\sffamily\bfseries\tiny
    },
    add/.style={
        draw=trunkstroke,
        fill=white,
        circle,
        thick,
        inner sep=2.5pt,
        font=\sffamily\bfseries\tiny,
        text=trunkstroke
    },
    rep/.style={
        minimum width=1.3cm,
        minimum height=2cm,
        inner sep=0pt
    },
    flow/.style={
        -{Stealth[length=5pt,width=4pt]},
        thick,
        draw=arrowcolor
    }
}

\newcommand{\repcard}[3]{%
    \pgfmathsetseed{#1}%

    \draw[card] (-0.65,-1) rectangle (0.65,1);

    \begin{scope}[shift={(-0.45,-0.07)}]
        \foreach \col in {0,...,4} {
            \foreach \row in {0,...,4} {
                \pgfmathsetmacro{\val}{int(random(#2,#3))}
                \fill[pairstroke!\val!white]
                    ({\col*0.18},{\row*0.18})
                    rectangle ++(0.18,0.18);
            }
        }

        \foreach \divider in {1,...,4} {
            \draw[white,line width=0.25pt]
                ({\divider*0.18},0) -- ++(0,0.9);
            \draw[white,line width=0.25pt]
                (0,{\divider*0.18}) -- ++(0.9,0);
        }

        \draw[very thick,pairstroke] (0,0) rectangle (0.9,0.9);
    \end{scope}

    \draw[cardborder!30,thin]
        (-0.45,-0.18) -- (0.45,-0.18);

    \begin{scope}[shift={(-0.45,-0.73)}]
        \foreach \col in {0,...,4} {
            \pgfmathsetmacro{\val}{int(random(#2,#3))}
            \fill[singlestroke!\val!white]
                ({\col*0.18},0)
                rectangle ++(0.18,0.3);
        }

        \foreach \divider in {1,...,4} {
            \draw[white,line width=0.25pt]
                ({\divider*0.18},0) -- ++(0,0.3);
        }

        \draw[very thick,singlestroke]
            (0,0) rectangle (0.9,0.3);
    \end{scope}
}

\begin{tikzpicture}[node distance=0.8cm and 0.8cm]
    \node (rep0) [rep] at (0,0) {};
    \begin{scope}[shift={(rep0.center)}]
        \repcard{101}{10}{35}
    \end{scope}

    \node (plus) [add,right=1.2cm of rep0] {$+$};

    \node (trunk) [rblock=trunk,right=0.8cm of plus]
        {Trunk};

    \node (rep1) [rep] at ($(trunk.east)+(1.2,0)$) {};
    \begin{scope}[shift={(rep1.center)}]
        \repcard{202}{25}{85}
    \end{scope}

    \node (critic) [rblock=critic,above=0.55cm of rep1]
        {CycleCritic};

    \node (gate) [gate,right=0.8cm of rep1]
        {\phantom{Accept?}};

    \node (d1) [decision] at (critic.east -| gate)
        {\textcolor{checkgreen}{\small$\checkmark$}\,/\,%
         \textcolor{crossred}{\small$\times$}\,/\,?};

    \path (rep1.south) ++(0,-0.35cm) coordinate (loopbottom);

    \begin{scope}[on background layer]
        \node (container) [
            draw=containerstroke,
            fill=containerbg,
            thick,
            rounded corners=14pt,
            fit={
                (plus)
                (trunk)
                (rep1)
                (critic)
                (d1)
                (gate)
                (loopbottom)
            },
            inner xsep=16pt,
            inner ysep=9pt
        ] {};
    \end{scope}
    
    \node (emb) [rblock=embed,above=0.2cm of rep0]
        {Input\\embedding};

    \node (seq) [
        above=0.13cm of emb,
        font=\ttfamily\bfseries\small,
        text=embedstroke!90!black,
        inner sep=1pt
    ] {Y\,C\,I\,P\,S};
    
    \node (seq_label) [
        above=0.05cm of seq,
        font=\sffamily\scriptsize,
        text=cardborder,
        inner sep=0pt
    ] {Input sequence};
    \node [
        anchor=north west,
        font=\sffamily\bfseries,
        text=containerstroke!90!black,
        xshift=6pt,
        yshift=-4pt
    ] at (container.north west)
        {Recycling Loop};

    \node (struct) [
        rblock=struct,
        right=1.2cm of container,
        yshift=1.4cm
    ] {Structure\\module};

    \node (conf) [rblock=conf,below=0.8cm of struct]
        {Confidence\\module};

    \node (dec_conf) [decision,below=0.6cm of conf]
        {\textcolor{checkgreen}{\small$\checkmark$}\,/\,%
         \textcolor{crossred}{\small$\times$}};

    \draw[flow] (seq.south) -- (emb.north);
    \draw[flow] (emb.south) -- (rep0.north);
    \draw[flow] (rep0.east) -- (plus.west);
    \draw[flow] (plus.east) -- (trunk.west);
    \draw[flow] (trunk.east) -- (rep1.west);
    \draw[flow] (rep1.north) -- (critic.south);
    \draw[flow] (critic.east) -- (d1.west);
    \draw[flow,dashed] (d1.south) -- (gate.north);
    \draw[flow] (rep1.east) -- (gate.west);

    \draw[flow,rounded corners=6pt]
        (gate.south)
        -- (gate.south |- loopbottom)
        -- (plus.south |- loopbottom)
        -- (plus.south);

    \coordinate (container_exit)
        at (container.east |- struct.west);

    \coordinate (split_pt)
        at ($(container_exit)!0.5!(struct.west)$);

    \draw[flow] (container_exit) -- (struct.west);
    \draw[flow,rounded corners=6pt] (split_pt) |- (conf.west);
    \draw[flow] (struct.south) -- (conf.north);
    \draw[flow] (conf.south) -- (dec_conf.north);
\end{tikzpicture}

%% file: CC_combined_arch.tex
\begin{figure}[htbp]
  \centering
  \resizebox{\textwidth}{!}{%
\begin{tikzpicture}[scale=0.7]
  \tikzstyle{sectionhead}=[font=\large\bfseries, align=center]
  \tikzstyle{badge}=[draw=gray!50, fill=shapeblue, rounded corners=3pt, inner sep=3pt, font=\small]
  \tikzstyle{badgeblue}=[draw=gray!50, fill=shapeblue, rounded corners=3pt, inner sep=3pt, font=\small]
  \colorlet{sccc}{histblue!70!black}          

  \node[sectionhead] at (0.5, 3.3) {per pair embedding};

  \begin{scope}[shift={(0,0)}]
    \draw[fill=blockgreen!80, draw=gridline, thick] (-1.5,-1.5) rectangle (1.5,1.5);
    \foreach \i in {1,...,5} {
      \draw[gridline] (-1.5 + \i*0.5, -1.5) -- (-1.5 + \i*0.5, 1.5);
      \draw[gridline] (-1.5, -1.5 + \i*0.5) -- (1.5, -1.5 + \i*0.5);
    }
    \draw[fill=blockgreen!50, draw=gridline, thick] (-1.5,1.5) -- (-0.7,2.3) -- (2.3,2.3) -- (1.5,1.5) -- cycle;
    \foreach \i in {1,...,5} {
      \draw[gridline] (-1.5 + \i*0.5, 1.5) -- (-0.7 + \i*0.5, 2.3);
    }
    \foreach \j in {1,...,3} {
      \draw[gridline] (-1.5 + \j*0.2, 1.5 + \j*0.2) -- (1.5 + \j*0.2, 1.5 + \j*0.2);
    }
    \draw[fill=blockgreen!90, draw=gridline, thick] (1.5,-1.5) -- (2.3,-0.7) -- (2.3,2.3) -- (1.5,1.5) -- cycle;
    \foreach \i in {1,...,5} {
      \draw[gridline] (1.5, -1.5 + \i*0.5) -- (2.3, -0.7 + \i*0.5);
    }
    \foreach \j in {1,...,3} {
      \draw[gridline] (1.5 + \j*0.2, -1.5 + \j*0.2) -- (1.5 + \j*0.2, 1.5 + \j*0.2);
    }

    \node[below] at (0, -1.6) {$n$};
    \node[left] at (-1.6, 0) {$n$};
    \node[below right] at (1.9, -1.0) {$d$};
    \node[badge] at (0.2, -2.5) {Shape: (n, n, $d$)};
  \end{scope}

  \draw[->, ultra thick] (2.7, 0) -- (4.0, 0);

  \begin{scope}[shift={(5.6,0)}]
    \draw[fill=gray!15, draw=black!80, thick] (-1.35, -1.9) -- (1.35, -2.8) -- (1.35, 2.8) -- (-1.35, 1.9) -- cycle;
    \node[align=center, font=\Large\bfseries] at (0,0) {Shared\\FFN};
  \end{scope}

  \draw[->, thick] (7.0, 0.7) -- (8.4, 2.0);
  \draw[->, thick] (7.0, -0.7) -- (8.4, -2.0);

  \begin{scope}[shift={(9.6,2.0)}]
    \draw[fill=blockgreen!80, draw=gridline, thick] (-1.2,-1.2) rectangle (1.2,1.2);
    \foreach \i in {1,...,5} {
      \draw[gridline] (-1.2 + \i*0.4, -1.2) -- (-1.2 + \i*0.4, 1.2);
      \draw[gridline] (-1.2, -1.2 + \i*0.4) -- (1.2, -1.2 + \i*0.4);
    }
    \draw[fill=blockgreen!50, draw=gridline, thick] (-1.2,1.2) -- (-0.2,2.0) -- (2.2,2.0) -- (1.2,1.2) -- cycle;
    \foreach \i in {1,...,5} {
      \draw[gridline] (-1.2 + \i*0.4, 1.2) -- (-0.2 + \i*0.4, 2.0);
    }
    \foreach \j in {1,...,4} {
      \draw[gridline] (-1.2 + \j*0.2, 1.2 + \j*0.16) -- (1.2 + \j*0.2, 1.2 + \j*0.16);
    }
    \draw[fill=blockgreen!90, draw=gridline, thick] (1.2,-1.2) -- (2.2,-0.4) -- (2.2,2.0) -- (1.2,1.2) -- cycle;
    \foreach \i in {1,...,5} {
      \draw[gridline] (1.2, -1.2 + \i*0.4) -- (2.2, -0.4 + \i*0.4);
    }
    \foreach \j in {1,...,4} {
      \draw[gridline] (1.2 + \j*0.2, -1.2 + \j*0.16) -- (1.2 + \j*0.2, 1.2 + \j*0.16);
    }

    \node[below right] at (1.7, -0.7) {$l$};
    \node[badge] at (0.2, -1.8) {Shape: (n, n, l)};
  \end{scope}

  \begin{scope}[shift={(9.6,-2.0)}]
    \draw[fill=blockgreen!80, draw=gridline, thick] (-1.2,-1.2) rectangle (1.2,1.2);
    \foreach \i in {1,...,5} {
      \draw[gridline] (-1.2 + \i*0.4, -1.2) -- (-1.2 + \i*0.4, 1.2);
      \draw[gridline] (-1.2, -1.2 + \i*0.4) -- (1.2, -1.2 + \i*0.4);
    }
    \node[below, font=\bfseries] at (0, -1.3) {Weight Matrix};
    \node[badge] at (0, -2.4) {Shape: (n, n)};
  \end{scope}

  \draw[->, thick] (11.8, 2.0) -- (12.5, 0.3);
  \draw[->, thick] (10.8, -2.0) -- (12.5, -0.3);

  \node[circle, draw=black!80, fill=white, inner sep=1pt, thick, font=\Large] (otimes) at (12.8, 0) {$\otimes$};

  \draw[->, ultra thick] (otimes) -- (14.2, 0);

  \begin{scope}[shift={(14.8,0)}]
    \node[above, font=\bfseries, align=center] at (0, 1.4) {Pooled\\Vector};
    \draw[fill=blockgreen!80, draw=gridline, thick] (-0.3,-1.25) rectangle (0.3,1.25);
    \foreach \i in {1,...,4} {
      \draw[gridline] (-0.3, -1.25 + \i*0.5) -- (0.3, -1.25 + \i*0.5);
    }
    \node[badge] at (0, -2.0) {Shape: (l,)};
  \end{scope}

  \draw[->, ultra thick] (15.4, 0) -- (16.4, 0);

  \begin{scope}[shift={(17.5,0)}]
    \draw[fill=gray!15, draw=black!80, thick] (-0.9, -1.6) -- (0.9, -2.4) -- (0.9, 2.4) -- (-0.9, 1.6) -- cycle;
    \node[align=center, font=\Large\bfseries] at (0,0) {FFN\\Head};
  \end{scope}

  \draw[->, ultra thick] (18.5, 0) -- (20.6, 0);

  \begin{scope}[shift={(22.0,0)}]
    \draw[dashed, rounded corners=6pt, fill=gray!5] (-1.3,-1.3) rectangle (1.3,1.3);
    \draw[fill=histblue, draw=black!70] (-1.1,-1.0) rectangle (-0.85,-0.5);
    \draw[fill=histblue, draw=black!70] (-0.85,-1.0) rectangle (-0.6,0.3);
    \draw[fill=histblue, draw=black!70] (-0.6,-1.0) rectangle (-0.35,0.8);
    \draw[fill=histblue, draw=black!70] (-0.35,-1.0) rectangle (-0.1,0.0);
    \draw[fill=histblue, draw=black!70] (-0.1,-1.0) rectangle (0.15,-0.3);
    \draw[fill=histblue, draw=black!70] (0.15,-1.0) rectangle (0.4,0.5);
    \draw[fill=histblue, draw=black!70] (0.4,-1.0) rectangle (0.65,-0.1);
    \draw[fill=histblue, draw=black!70] (0.65,-1.0) rectangle (0.9,-0.7);
    \draw[fill=histblue, draw=black!70] (0.9,-1.0) rectangle (1.15,-0.9);
    \node[below, font=\small, align=center] at (0, -1.5) {predicted metric\\distribution};
  \end{scope}

\end{tikzpicture}
  }
  \caption{\textbf{Schematic architecture of the learned CycleCritics.} Both apply a shared feedforward
  network to a latent representation per pair representation \emph{individually} before aggregating by weighted
  mean pooling. The structure critic additionally receives the pair distances of the
  designed backbone; the predicted metric can be e.g. ipSAE or ipTM for the confidence critic and
  self-consistency RMSD for the structure critic.}
  \label{fig:cyclecritic-combined}
\end{figure}

%% file: sccc_architecture.tex
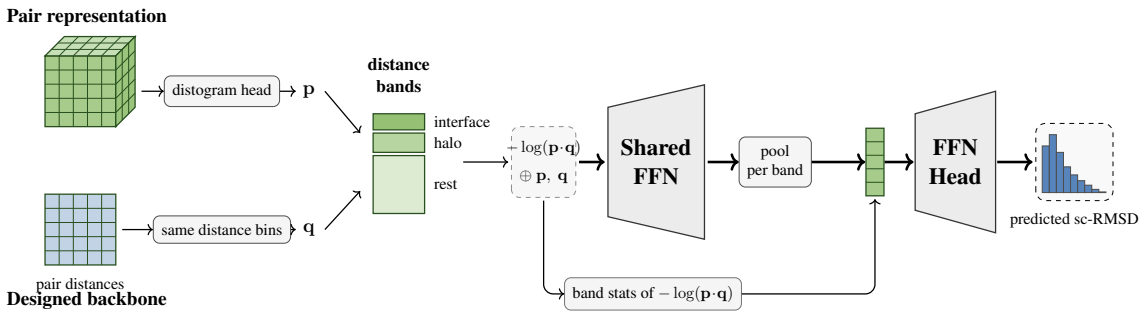
\begin{figure}[htbp]
  \centering
  \resizebox{0.92\textwidth}{!}{%
\begin{tikzpicture}[scale=0.7]
  \tikzstyle{sectionhead}=[font=\large\bfseries, align=center]
  \tikzstyle{badge}=[draw=gray!50, fill=badgegreen, rounded corners=3pt, inner sep=3pt, font=\small]
  \tikzstyle{op}=[draw=black!70, fill=gray!8, rounded corners=4pt, align=center, font=\small, inner sep=5pt]
  \colorlet{blockref}{histblue!45}

  \node[sectionhead] at (0.2, 4.2) {Pair representation};
  \begin{scope}[shift={(0,1.9)}]
    \draw[fill=blockgreen!80, draw=gridline, thick] (-1.1,-1.1) rectangle (1.1,1.1);
    \foreach \i in {1,...,4} {
      \draw[gridline] (-1.1 + \i*0.44, -1.1) -- (-1.1 + \i*0.44, 1.1);
      \draw[gridline] (-1.1, -1.1 + \i*0.44) -- (1.1, -1.1 + \i*0.44); }
    \draw[fill=blockgreen!50, draw=gridline, thick] (-1.1,1.1) -- (-0.5,1.7) -- (1.7,1.7) -- (1.1,1.1) -- cycle;
    \foreach \i in {1,...,4} { \draw[gridline] (-1.1 + \i*0.44, 1.1) -- (-0.5 + \i*0.44, 1.7); }
    \foreach \j in {1,2} { \draw[gridline] (-1.1 + \j*0.2, 1.1 + \j*0.2) -- (1.1 + \j*0.2, 1.1 + \j*0.2); }
    \draw[fill=blockgreen!90, draw=gridline, thick] (1.1,-1.1) -- (1.7,-0.5) -- (1.7,1.7) -- (1.1,1.1) -- cycle;
    \foreach \i in {1,...,4} { \draw[gridline] (1.1, -1.1 + \i*0.44) -- (1.7, -0.5 + \i*0.44); }
    \foreach \j in {1,2} { \draw[gridline] (1.1 + \j*0.2, -1.1 + \j*0.2) -- (1.1 + \j*0.2, 1.1 + \j*0.2); }
  \end{scope}
  \node[op] (dh) at (4.4, 1.9) {distogram head};
  \draw[->, thick] (1.9,1.9) -- (dh.west);
  \draw[->, thick] (dh.east) -- (6.7,1.9);
  \node[font=\large] at (7.1,1.9) {$\mathbf{p}$};

  \node[sectionhead] at (0.2,-4.6) {Designed backbone};
  \begin{scope}[shift={(0,-2.4)}]
    \draw[fill=blockref!80, draw=gridline, thick] (-1.1,-1.1) rectangle (1.1,1.1);
    \foreach \i in {1,...,4} {
      \draw[gridline] (-1.1 + \i*0.44, -1.1) -- (-1.1 + \i*0.44, 1.1);
      \draw[gridline] (-1.1, -1.1 + \i*0.44) -- (1.1, -1.1 + \i*0.44); }
    \node[below, font=\small] at (0,-1.3) {pair distances};
  \end{scope}
  \node[op] (sb) at (4.4,-2.4) {same distance bins};
  \draw[->, thick] (1.3,-2.4) -- (sb.west);
  \draw[->, thick] (sb.east) -- (6.7,-2.4);
  \node[font=\large] at (7.1,-2.4) {$\mathbf{q}$};

  \draw[->, thick] (7.6,1.9) -- (8.7,0.6);
  \draw[->, thick] (7.6,-2.4) -- (8.7,-1.2);
  \begin{scope}[shift={(9.9,-0.3)}]
    \node[sectionhead, font=\bfseries, align=center] at (0,2.7) {distance\\bands};
    \draw[fill=blockgreen!95, draw=gridline, thick] (-0.8,1.0) rectangle (0.8,1.5);
    \draw[fill=blockgreen!65, draw=gridline, thick] (-0.8,0.3) rectangle (0.8,0.9);
    \draw[fill=blockgreen!30, draw=gridline, thick] (-0.8,-1.6) rectangle (0.8,0.2);
    \node[right, font=\small] at (0.9,1.25) {interface};
    \node[right, font=\small] at (0.9,0.6) {halo};
    \node[right, font=\small] at (0.9,-0.7) {rest};
  \end{scope}

  \draw[->, thick] (11.9,-0.3) -- (13.3,-0.3);

  \begin{scope}[shift={(14.4,-0.3)}]
    \draw[dashed, rounded corners=5pt, fill=gray!6, draw=black!45] (-1.0,-1.1) rectangle (1.0,1.1);
    \node[font=\small, align=center] at (0,0.45) {$-\log(\mathbf{p}\!\cdot\!\mathbf{q})$};
    \node[font=\small, align=center] at (0,-0.45) {$\oplus\ \mathbf{p},\ \mathbf{q}$};
  \end{scope}

  \draw[->, ultra thick] (15.5,-0.3) -- (16.3,-0.3);

  \begin{scope}[shift={(17.9,-0.3)}]
    \draw[fill=gray!15, draw=black!80, thick] (-1.5,-1.7) -- (1.5,-2.4) -- (1.5,2.4) -- (-1.5,1.7) -- cycle;
    \node[align=center, font=\Large\bfseries] at (0,0) {Shared\\FFN};
  \end{scope}

  \node[op] (pool) at (21.6,-0.3) {pool\\per band};
  \draw[->, ultra thick] (19.5,-0.3) -- (pool.west);

  \node[op, draw=black!55] (bs) at (17.8,-4.4) {band stats of $-\log(\mathbf{p}\!\cdot\!\mathbf{q})$};
  \draw[->, thick, rounded corners=6pt] (14.4,-1.5) -- (14.4,-4.4) -- (bs.west);
  \draw[->, thick, rounded corners=6pt] (bs.east) -- (24.7,-4.4) -- (24.7,-1.45);

  \draw[->, ultra thick] (pool.east) -- (24.4,-0.3);

  \begin{scope}[shift={(24.7,-0.3)}]
    \draw[fill=blockgreen!80, draw=gridline, thick] (-0.28,-1.05) rectangle (0.28,1.05);
    \foreach \i in {1,...,4} { \draw[gridline] (-0.28,-1.05 + \i*0.42) -- (0.28,-1.05 + \i*0.42); }
  \end{scope}

  \draw[->, ultra thick] (25.0,-0.3) -- (25.9,-0.3);

  \begin{scope}[shift={(27.2,-0.3)}]
    \draw[fill=gray!15, draw=black!80, thick] (-1.25,-1.5) -- (1.25,-2.2) -- (1.25,2.2) -- (-1.25,1.5) -- cycle;
    \node[align=center, font=\Large\bfseries] at (0,0) {FFN\\Head};
  \end{scope}

  \draw[->, ultra thick] (28.5,-0.3) -- (29.6,-0.3);

  \begin{scope}[shift={(30.9,-0.3)}]
    \draw[dashed, rounded corners=6pt, fill=gray!5] (-1.2,-1.2) rectangle (1.2,1.2);
    \foreach \x/\h in {-1.0/0.5,-0.78/0.85,-0.56/0.3,-0.34/-0.15,-0.12/-0.4,0.1/-0.58,0.32/-0.72,0.54/-0.85,0.76/-0.93} {
      \draw[fill=histblue, draw=black!70] (\x,-0.95) rectangle (\x+0.22,\h); }
    \node[below, font=\small, align=center] at (0,-1.4) {predicted sc-RMSD};
  \end{scope}
\end{tikzpicture}
  }
  \caption{\textbf{Structure CycleCritic.} The trunk's predicted distogram $\mathbf{p}$ and
  the designed backbone's pair distances $\mathbf{q}$ are encoded on the same distance bins,
  so their per-pair agreement $-\log(\mathbf{p}\!\cdot\!\mathbf{q})$ is the analytic gate.
  Pairs are banded by distance --- interface ($<12$\,\AA), halo ($12$--$18$\,\AA) and a
  subsample of the rest --- and pooled within bands. Band statistics of the agreement
  bypass the shared network, so the analytic gate stays a linear readout. The head predicts
  the distribution of self-consistency RMSD.}
  \label{fig:sccc-architecture}
\end{figure}